# Direct measurement of the $^{23}$Mg($p$, $\gamma$)$^{24}$Al reaction using the DRAGON recoil separator and a hybrid array of $\gamma$-ray detectors

A.C. Edwin[1,*], G. Christian[1,†], A. Lennarz[2,3], I. Casandjian[3], S.M. Collins[4,5], E. Desmarais[1], U. Greife[6], D. A. Hutcheon[2], A. Katrusiak[2,a], M. Loria[2,7], S. Molló[2,8], J. O'Connell[1], S. Pascu[5,9], L. Pedro-Botet[2,8], Zs. Podolyák[5], B. J. Reed[5,b], P. H. Regan[4,5], C. Ruiz[2,7], R. Shearman[4], S. Upadhyayula[2,c], L. Wagner[2,d] and M. Williams[5]

[1]Department of Astronomy & Physics, Saint Mary's University, Halifax, NS B3H 3C3, Canada
[2]TRIUMF, 4004 Wesbrook Mall, Vancouver, BC V6T 2A3, Canada
[3]Department of Physics & Astronomy, McMaster University, Hamilton, ON L8S 4M1, Canada
[4]National Physical Laboratory, Teddington, TW11 0LW, UK
[5]School of Mathematics & Physics, University of Surrey, Guildford, GU2 7XH, UK
[6]Department of Physics, Colorado School of Mines, Golden, Colorado 80401, USA
[7]Department of Physics & Astronomy, University of Victoria, Victoria, BC V8W 3P6, Canada
[8]Departament de Física, Universitat Politècnica de Catalunya, Barcelona, Spain
[9]National Institute for Physics and Nuclear Engineering, R-77125, Bucharest-Magurele, Romania

## ABSTRACT

**Background:** The $^{23}$Mg$(p,\gamma)^{24}$Al reaction plays a key role in nova nucleosynthesis, serving as a bridge between the Ne-Na and Mg-Al cycles and impacting synthesis of a wide variety of nuclides. It is crucial that the rate of this reaction be well constrained by experimental data, in particular the properties of the key resonance at $E_{cm} \approx 480$ keV, which dominates the rate at nova temperatures.

**Purpose:** Our study aims to improve experimental constraints on the energy and strength of the key $E_{cm} \approx 480$ keV $^{23}$Mg$(p,\gamma)^{24}$Al resonance through a new direct measurement of its strength and energy.

**Methods:** We used a $^{23}$Mg radioactive beam to directly observe capture into the $E_{cm} \approx 480$ keV resonance in inverse kinematics. The $^{24}$Al recoils were selected by the DRAGON recoil separator, and coincident $\gamma$ rays were measured in a hybrid LaBr$_3$/CeBr$_3$/BGO scintillator array. The $^{24}$Al singles yield was used to determine the resonance strength, while the time difference between the coincident $\gamma$ rays and the incident beam was used to determine the resonance energy.

**Results**: We obtained a strength of $\omega\gamma = 38(7)$ meV and an energy of $E_{cm} = 478.8(26)$ keV for the key astrophysical $^{23}$Mg(p, $\gamma$)$^{24}$Al resonance.

**Conclusions:** Our reported strength is in excellent agreement with the previously reported direct measurement, as well as shell model calculations. Our energy is in good agreement with earlier indirect measurements. Reaction rate calculations incorporating our results indicate that the $^{23}$Mg$(p,\gamma)^{24}$Al rate is constrained to a $1\sigma$ up/down factor of 1.5 or smaller in the nova temperature range of 0.2–0.4 GK.

## I. INTRODUCTION

Classical novae are stellar explosions resulting from the accretion of hydrogen-rich material onto a white dwarf. They are responsible for the formation of a significant number of nuclides lighter than calcium. Sensitivity studies, which model the rate of nuclear reaction variations on nucleosynthesis predictions, have determined a substantial number of proton capture reactions whose rates have an observable effect on nova nucleosynthesis [1]. One of these reactions is the proton capture on radioactive $^{23}$Mg, resulting in $\gamma$-ray emission from $^{24}$Al – the $^{23}$Mg$(p,\gamma)^{24}$Al reaction. This reaction has an influence on the synthesis of many nuclides including the long-lived radioactive isotopes $^{26}$Al ($t_{1/2} = 7.2 \times 10^5$ years) and $^{22}$Na ($t_{1/2} = 2.6$ years), which emit distinct $\gamma$ rays of energies 1809 keV and 1275 keV, respectively, and are candidates for observation in $\gamma$-ray telescopes. The primary reason for this reaction's strong impact on nucleosynthesis is its role in connecting the Ne-Na and Mg-Al cycles, shown in Figure 1. As such, the reaction serves as a bridge between two distinct regions of the reaction network, impacting production rates both above and below Mg.

The $^{23}$Mg $(p,\gamma)^{24}$Al reaction has been studied extensively, both experimentally and theoretically. A variety of $^{24}$Mg($^3$He$,t)^{24}$Al experiments have identified resonances above the proton threshold in $^{24}$Al, measuring their excitation energies [2, 3, 4, 5]. These experiments established that the $J^\pi = 3^+$ resonance

*Contact author: Edwin.AmandaCrystal@smu.ca
†Contact author: Greg.Christian@smu.ca
[a]Present address: Department of Physics & Astronomy, University of Victoria, Victoria, BC V8W 3P6, Canada
[b]Present address: TRIUMF, 4004 Wesbrook Mall, Vancouver, BC V6T 2A3, Canada and Department of Astronomy & Physics, Saint Mary's University, Halifax, NS B3H 3C3, Canada
[c]Present address: Lawrence Livermore National Laboratory, Livermore, CA 94550, USA
[d]Present address: Facility for Rare Isotope Beams, Michigan State University, East Lansing, MI 48824, USA

around $E_x = 2345$ keV dominates the rate at nova temperatures. The $\gamma$-ray decay of this key resonance was observed in a fusion-evaporation experiment by Lotay *et al.* [6]. This experiment determined the excitation energy from the $\gamma$-ray energies, resulting in the most precise value to date, $E_x = 2345.1(14)$ keV. The experiment also firmly established the $3^+$ spin assignment. Most recently, a $^{24}$Mg($^3$He, $t$)$^{24}$Al experiment which also detected the protons from $^{24}$Al$^*$ decay, reported an excitation energy of 2342(4) keV for the key resonance and set a lower limit on its proton decay branch of $\Gamma_p/\Gamma_{\text{tot}} > 0.64$ [7].

Translation of the excitation energy into center-of-mass resonance energy requires knowledge of the proton separation threshold in $^{24}$Al. At the time of the Lotay *et al.* experiment, this was taken to be $S_p = 1872.15(307)$ keV, from the 2003 Atomic Mass Evaluation (AME-03) [8]. This resulted in a center-of-mass energy of $E_{cm} = 472.9(34)$ keV. The separation energy was revised by nearly 10 keV, to 1862.6(29) keV, by Erikson *et al.*, who re-calibrated the $^{24}$Mg$(p,n)^{24}$Al threshold measurement that had earlier been used to determine the $^{24}$Al mass [9]. The threshold was revised again to 1864.32(28) keV, following a Penning-trap measurement of the $^{24}$Al mass by Chowdhury *et al.* [10]. Most recently, the 2020 AME (AME-20) reported $S_p = 1864.11(23)$ keV, with the small change relative to Ref. [10] primarily driven by a revision of the $^{23}$Mg mass. Using the AME-20 threshold, together with the Lotay *et al.* excitation energy, results in $E_{cm} = 480.1(14)$ keV for the key $J^\pi = 3^+$ resonance.

Many of the indirect measurements, as well as the shell-model study by Herndl *et al.* [11], made attempts to determine the strength of the key $E_{cm} \approx 480$ keV resonance. These are mostly based on shell model calculations of the radiative and proton widths for the state, or on assumed analog assignments from the mirror state in $^{24}$Na. The resonance strength reported in most of these studies is around 25 meV (a comprehensive overview given in Section IV). The recent ($^3$He, $t$) measurement by Vyfers *et al.* reported a resonance strength of 19(9) meV, based on their shell-model calculation of $\Gamma_\gamma = 24$ meV, together with a Monte Carlo analysis of their observed ($^3$He, $t+p$) spectra [7]. In addition to the studies already mentioned, the reaction rate compilation of Iliadis *et al.* [12] adopted a proton width of $\Gamma_p = 170(40)$ meV for this level, based on the $L = 2$ spectroscopic factor reported for the mirror state in a $^{23}$Na$(d,p)^{24}$Na experiment [13].

A direct measurement of the key $E_{cm} \approx 480$ keV resonance was reported by Erikson *et al.* [9]. This experiment used the same DRAGON recoil separator at TRIUMF that is used in the current work [14]. The Erikson experiment was performed in inverse kinematics using a $^{23}$Mg radioactive beam, which suffered from significant $^{23}$Na isobaric contamination. The beam energy was $E/A = 502$ keV, chosen to center the expected (at the time of the experiment) $E_{cm} = 473$ keV resonance in the DRAGON extended gas target. Despite numerous challenges, the authors clearly observed $^{24}$Al + $\gamma$ ray coincidences resulting from the $^{23}$Mg$(p,\gamma)^{24}$Al reaction. An analysis of the $z$-position spectrum in their Bismuth Germanate (BGO) $\gamma$-ray detectors indicated that the resonant reaction occurred far upstream in the gas target, where the gas-density profile may not be uniform. A likelihood analysis of simulated resonance strength vs. resonance energy produced a bi-modal distribution, with significant covariance between the two parameters. The authors extracted resonance strength and energy values (with statistical uncertainties) by taking the mean and 16%/84% percentiles of the likelihood surface, projected separately onto the energy and strength axes. This resulted in a resonance strength of $\omega\gamma = 38(^{+21}_{-15})$ meV and an energy of $E_{cm} = 485.7(^{+13}_{-18})$ keV. The energy value was later revised to $485.0(^{+13}_{-18})$ keV based on a re-calibration of DRAGON's magnetic constant, which is used to measure the incident beam energy [10, 15]. A revised resonance strength of $26.6(^{+154}_{-70})$ meV was also reported by Wrede *et al.*, who folded the Erikson likelihood surface together with a probability density function dictated by the indirect measurements of the resonance energy [16]. This effectively amounts to extracting the resonance strength from the highest mode of the likelihood surface rather than the mean. While the reported strength by Erikson *et al.* is generally in good agreement with the indirect estimates, the $485.7(^{+13}_{-18})$ keV resonance energy quoted in the paper differs by $2.1\sigma$ from the most precise indirect measurement of 480.1(14) keV. With the revision to $485.0(^{+13}_{-18})$ keV, there is still a modest $1.7\sigma$ tension between the two measurements. This discrepancy, together with the target-density and beam-contamination issues encountered during the Erikson experiment, motivates a new direct measurement of the key $E_{cm} \approx 480$ keV resonance in the $^{23}$Mg$(p,\gamma)^{24}$Al reaction.

In this article, we report the results of a new direct measurement of the $^{23}$Mg$(p,\gamma)^{24}$Al reaction, using the DRAGON recoil separator coupled to a hybrid array of $LaBr_3$, $CeBr_3$, and BGO scintillators for prompt $\gamma$-ray detection [17]. The hybrid array was chosen for $\gamma$-ray detection to maximize timing resolution while using the new resonance timing technique to measure the energy of the $E_{cm} \approx 480$ keV resonance [17]. We also extracted a new strength value for this resonance from the singles yield of $^{24}$Al recoils. The present experiment uses a $^{23}$Mg radioactive beam of higher energy, $E/A$ = 507.6(10) keV, than the Erikson experiment. This places the resonance further downstream in the DRAGON gas target, ensuring that the full resonance is contained within the constant-pressure region of the target. The present experiment also uses a $^{23}$Mg radioactive beam with significantly improved purity as compared to the earlier Erikson measurement.

## II. EXPERIMENT

The experiment was performed at TRIUMF, Canada's particle accelerator centre. A $^{23}$Mg beam of energy $E/A$ = 507.6(10) keV and average intensity 2.3(3) $\times$ $10^7$s$^{-1}$ was produced at the TRIUMF ISAC-I facility using a 520 MeV proton-beam driver impinged on a SiC production target [18]. The beam was ionized to a $1^+$ charge state using the TRIUMF Resonant Ionization Laser Ion Source (TRILIS), which provides element-selective ionization based on atomic resonant absorption lines (in our case, selecting Mg), together with a background from surface-ionized isobars (in our case, dominated by stable $^{23}$Na). The TRILIS lasers are pulsed, giving the laser-ionized $^{23}$Mg a periodic time structure, while the surface-ionized $^{23}$Na is continuous in time. This difference was exploited by an RF kicker device synchronized with the laser pulse structure to preferentially transmit the laser-ionized beam, significantly reducing the $^{23}$Na contamination [19]. This resulted in an observed Mg:Na ratio, delivered to DRAGON, ranging from approximately 6:1 to 94:1, with no other $A = 23$ isobars present in measurable quantities. In contrast, the earlier Erikson *et al.* measurement, which did not use a kicker, was performed with an Mg:Na ratio ranging from approximately 1:20 to as low as approximately 1:1000 at various points during the beam time [9].

The beam was accelerated to its final energy using the ISAC-I radiofrequency quadrupole (RFQ) and drift-tube linac (DTL) accelerators. A periodic timing signal with frequency 11.79 MHz was taken from the RFQ and incorporated into the data acquisition system for use in the $\gamma$-ray timing analysis. The beam impinged onto the DRAGON windowless gas target, which was filled with $H_2$ gas maintained at a pressure of 10.5(2) mbar and temperature of 307.5(11) Kelvin. The charge state delivered to the experiment was $5^+$. The $^{24}$Al$^{(6+)}$ recoils resulting from the $^{23}$Mg$(p,\gamma)^{24}$Al reaction were transmitted to the end of DRAGON where they were detected in a pair of micro-channel plate detectors (MCPs), providing a local time of flight (TOF) measurement, along with a segmented ionization chamber (IC), providing energy loss signals across four anodes [20, 21]. The geometry of the MCP elements was different from the traditional geometry presented in Ref. [21]. The new geometry used in the present experiment is described in Figure 2. The IC was operated at a pressure of approximately 9.1 mbar and used a mylar entrance window of 0.9 $\mu$m thickness. Together, the IC and MCP signals provided sufficient information to cleanly separate the $^{24}$Al recoils from background. A schematic of the DRAGON system is shown in Figure 3.

The gas target was surrounded by a hybrid $LaBr_3/CeBr_3$/BGO $\gamma$-ray detector array which recorded prompt $\gamma$ rays from the $^{23}$Mg$(p,\gamma)^{24}$Al reaction [17]. A photograph of these detectors is shown in Figure 4. Due to background from 440 keV $\gamma$ rays emitted by $^{23}$Mg implanted on the target entrance aperture, the detectors were operated with high trigger thresholds, which were tuned online to reduce the data acquisition dead time to an acceptable level. This resulted in a $\gamma$-ray detection efficiency of approximately 3%, giving a yield of only two reliably identified recoil-$\gamma$ coincidences. However, as discussed in Section III.D, this was sufficient to extract an energy for the $E_{cm} \approx$ 480 keV resonance using the resonance timing technique. While the low $\gamma$-ray yield prevented us from extracting a resonance strength from $^{24}$Al + $\gamma$-ray coincidences, the $^{24}$Al recoils were cleanly separated in heavy-ion singles. This allowed us to determine the resonance strength using the singles yield instead.

Unreacted $^{23}$Mg$^{(6+)}$ beam particles were deposited on slits just downstream of DRAGON's first electrostatic dipole. Viewing these slits was a decay station dedicated to determining the incident rate of the $^{23}$Mg beam. A fraction of the $\beta^+$ particles emitted by $^{23}$Mg decay ($t_{1/2} = 11.3$ s) entered a stainless-steel "horn" which provided material for $e^+$-$e^-$ annihilation. This resulted in emission of coincident 511 keV $\gamma$ rays from the horn, which were detected using a pair of NaI scintillators located at 90° relative to the long axis of the horn. A single high-purity germanium (HPGe) detector was also placed near the mass slits and used to detect the 440 keV $\gamma$-rays emitted by $^{23}$Mg decay with a 7.85% branching fraction [22]. In addition to the decay station detectors, the total beam current was monitored approximately every hour using a series of Faraday cups located just upstream and just downstream of the target, as well as just downstream of the charge slits following the first magnetic dipole. The Mg:Na ratio of the $A = 23$ beam was also monitored periodically as explained in Section III.A. Together, these detectors provided a reliable measure of the incident $^{23}$Mg beam rate.

## III. DATA ANALYSIS

### A. Beam Composition & Rate

To characterize its Mg:Na ratio, the beam was periodically attenuated using a "pepper-pot" device and directed into the IC, where the $^{23}$Mg and $^{23}$Na components could be separated based on energy loss signals. The attenuation factor was around $10^{-6}$, although this can vary due to slightly different attenuator positions between insertions, or due to small shifts in the beam profile. However, since the beam profile was insensitive to the species, the Mg and Na attenuation factors were identical. A 2D histogram of energy loss in the IC's second-most upstream anode ("Anode 1") vs. the sum of all anodes provided the best separation between the $^{23}$Mg and $^{23}$Na components of the beam. The identification of the $^{23}$Mg and $^{23}$Na components of the total spectrum was achieved by blocking the TRILIS ionization laser, which significantly enhanced the $^{23}$Na component of the beam. This can be seen in Figure 5, which shows the attenuated-beam IC spectra with TRILIS both on and off. In the upper panel, TRILIS off, the locus of events in the upper-right is clearly enhanced, confirming its identity as $^{23}$Na, while the lower panel, TRILIS on, shows an enhancement of the $^{23}$Mg events in the lower-left locus. Similar attenuated beam runs were taken periodically throughout the experiment, and for each run the ratio of Mg:Na isobars, with TRILIS on, was determined. These ratios are listed in Table 1. The Mg:Na ratio varied between a maximum of 94.4:1 and a minimum of 6.61:1 throughout the experiment. We noted no systematic reason for the variations, and the ratio appears to fluctuate randomly.

The rate of incident $^{23}$Mg ions was determined independently using both the NaI and HPGe detector systems near the mass slits. Energy spectra from both detectors are shown in Figure 6. The spectrum of coincidence signals from the NaI detectors (panel a) shows a clear locus around channel 1200-1400 on both axes, consisting of 511 keV events. The location of the 511 keV annihilation $\gamma$ rays in the uncalibrated spectrum was confirmed using a $^{22}$Na source. Likewise, the HPGe energy spectrum (panel b) clearly shows the 440 keV $\gamma$-ray peak from $^{23}$Mg decay, as well as a background peak from the 511 keV $\gamma$ rays. The location of these peaks in the uncalibrated energy spectrum was identified using a $^{152}$Eu source. For each of these detector systems, the integral of the peak of interest was determined for each production run. With appropriate corrections for the detector efficiencies, these give the total integrated $^{23}$Mg beam intensity on target.

The efficiency correction for the NaI and HPGe systems was performed by normalizing to periodic data taken throughout the experiment where the incident $^{23}$Mg rate was well known. Each of these normalization runs consisted of 300 s of data taking, immediately following an attenuated beam run characterizing the Mg:Na ratio, as well as a set of Faraday cup readings measuring the total incident beam current. Hence, for these normalization runs, the $^{23}$Mg rate was well known and could be used to determine a normalization factor, $\rho$, relating the measured rate in the decay-station detectors to the true incident beam rate:

$$R_{\mathrm{Mg}} = R_d/(\eta_{\mathrm{live}}\rho), \tag{1}$$

where $R_{\mathrm{Mg}}$ is the incident $^{23}$Mg rate (calculated from the Faraday cup), $R_d$ is the decay-station detector rate (coincident NaI or HPGe 440 keV peak), and $\eta_{\mathrm{live}}$ is the data acquisition live time, which cancels in the final resonance-strength determination since the same live time correction is needed in the recoil yield analysis. For each 300 s normalization run, $\rho$ can be determined by inverting Equation 1. This requires knowledge of $R_{\mathrm{Mg}}$, which we obtained from the current measurement on the Faraday cup just downstream of the gas target ("FC1"), together with the measured Mg:Na ratio from the preceding attenuated beam run. The FC1 cup views the beam in a mixture of charge states exiting the target, as well as a mixture of Mg and Na species. As a result, the incident $^{23}$Mg rate is given by

$$R_{\mathrm{Mg}} = \frac{I}{Q_e}\left(\sum_{q=1}^{12} q f_q^{(\mathrm{Mg})} + \frac{1}{r}\sum_{q=1}^{11} q f_q^{(\mathrm{Na})}\right), \tag{2}$$

where $I$ is the measured current on FC1, $Q_e$ is the electron charge, $r$ is the Mg:Na ratio from the preceding attenuated beam run, and $f_q^{(\mathrm{Mg})}$ and $f_q^{(\mathrm{Na})}$ are the respective Mg and Na charge state fractions for charge $q$. These charge state fractions were taken from Gaussian fits to the data reported in Ref. [23], interpolated to the present beam energy. An advantage of using FC1 for the normalization, rather than the target-upstream cup ("FC4") is that it implicitly includes the target transmission factor on a run-by-run basis. When inserted, both cups detect 100% of the incident beam, as their diameters (1″ FC4, 2″ FC1) are substantially larger than the beam spot size (approximately 2 mm).

In total, six separate normalization runs were performed, and the resulting $\rho$ values are shown in Figure 7, for both detector systems. The variation in $\rho$ values was consistent with a Gaussian-distributed random variable, and hence an overall normalization factor, with uncertainty, for each system was determined from the mean and standard deviation of the individual values. The resulting $\rho$ values are $9.8(15) \times 10^{-8}$ and $3.7(7) \times 10^{-8}$ for the NaI and HPGe systems, respectively. Using these normalization factors, separate integrated $^{23}$Mg rates were obtained using both systems. The resulting beam rates, on a run-by-run basis are shown in Figure 8. As seen in the figure, rates obtained using the different systems were consistent within uncertainties. The resulting total number of delivered $^{23}$Mg particles was $6.5(10) \times 10^{12}$ and $6.3(11) \times 10^{12}$ for the NaI and HPGe systems, respectively. These correspond to average rates of $2.3(4) \times 10^{7}$ s$^{-1}$ and $2.2(4) \times 10^7$ s$^{-1}$. For the final yield analysis, we took the weighted average of the two systems, resulting in a total of $6.4(8) \times 10^{12}$ $^{23}$Mg particles and an average $^{23}$Mg rate of $2.3(3) \times 10^7$ s$^{-1}$. These numbers represent beam transmitted through the DRAGON gas target, i.e. they include the empty-target target transmission factor that is often quoted as a detection efficiency in DRAGON experiments. The total number of $^{23}$Mg particles given above, $6.4(8) \times 10^{12}$, includes a correction for the 73.5% data-acquisition live time. In the resonance strength analysis (Section III.C and Table 2), the live time correction was omitted, which gives a total particle number of $4.7(6) \times 10^{12}$. The reason for the live-time omission is that the same correction applies to the recoil yield, and hence, it cancels in the final yield calculation.

### B. MCP and IC Efficiencies

As explained in Section III.C, the measured recoil yield must be corrected for the efficiencies of the MCP and IC end detectors. The total efficiency of these detectors is the product of the individual detection efficiencies for the MCPs and the IC ($\eta_{MCP}$ and $\eta_{IC}$, respectively), as well as the transmission through the various MCP foils and wire planes, $\tau_{MCP}$. The MCP transmission, for the present geometry pictured in Figure 2, was measured in five independent off-line measurements using an $\alpha$ source placed upstream of the MCPs. For these measurements, a double-sided silicon strip detector (DSSSD) was placed downstream of the MCPs to detect the $\alpha$ particles. In this setup, the MCP transmission is simply the ratio of MCP-inserted to MCP-retracted counting rates observed in the DSSSD. A sixth transmission measurement was also taken during a different experiment, which measured a high-yield resonance in the $^{15}$N($\alpha,\gamma$)$^{19}$F reaction. This experiment measured the recoil yield with the MCPs both inserted and retracted, and the transmission was taken as the ratio of inserted:retracted recoil yields. The transmission values obtained from the various source measurements, as well as the $^{15}$N(α,γ)$^{19}$F measurement, were consistent within uncertainties. Their weighted average of $\tau_{mcp} = 50.2(11)\%$ is used for the present analysis. A separate measurement was also taken, using the off-line $\alpha$ source, of the combined transmission of all wire planes associated with the downstream MCP (WP 4, 5, and 6 in Figure 2). This combined transmission, $\tau_{wire1}$, is relevant for extracting the IC efficiency and was measured to be $\tau_{wire1} = 72.3(11)\%$.

The MCP and IC detection efficiencies were extracted from attenuated-beam data taken during the present beam time, comparing the ratio of coincidence to singles count rates observed in each detector. In the present MCP configuration (Figure 2), beam ions that successfully pass through grid-1 will produce a detectable signal in MCP-1, even if they are later stopped in WP-4, 5, or 6. With this understanding, the detection efficiencies of both the MCP and IC can be extracted from attenuated beam data as follows:

$$\eta_{IC} = N_{both}/(N_{MCP}\tau_{wire1}) \quad (3)$$

$$\eta_{MCP} = N_{both}/N_{IC}, \quad (4)$$

where $N_{MCP}, N_{IC}$, and $N_{both}$ represent the number of attenuated beam counts in MCP TOF, IC energy loss, and MCP-IC coincidences, respectively, and $\tau_{wires1}$ is the combined WP-4, 5, and 6 transmission described previously. The resulting MCP and IC efficiencies were found to be 97.4(4)% and 96.0(15)%, respectively.

### C. Recoil ID & Resonance Strength

The $^{24}$Al recoils resulting from the $^{23}$Mg$(p,\gamma)^{24}$Al reaction were identified using a combination of signals from the MCPs, IC, and $\gamma$-ray detectors, with separate selections made for both heavy-ion singles and for heavy-ion + $\gamma$-ray coincidences. The recoil identification is summarized in Figure 9. Panel (a) of this figure shows a scatter plot of energy deposited in the second-most upstream anode of the IC ("anode 1") vs. the TOF between the two MCP detectors. The filled red circles represent heavy-ion singles events. Two clusters appear in this plot, with the lower-right cluster (circled in the figure) attributed to $^{24}$Al recoils and the upper-left cluster attributed to background from the $^{23}$Mg beam. The $^{24}$Al cluster contains a total of 63 events. The identity of the $^{24}$Al locus was based on the expected lower velocity of the recoils, as compared to the beam, resulting in a larger MCP TOF. Furthermore, the mean MCP TOF of the lower-right locus, approximately 54 ns, is consistent with calculations of the recoil velocity after the target, when accounting for timing offset corrections. The IC spectrum shows no evidence for $^{24}$Mg recoils resulting from $(p,\gamma)$ reactions on the contaminant $^{23}$Na beam. These events would be expected to display a different energy loss in anode 1 than the $^{24}$Al recoils.

From the 63 identified singles recoils, we extracted a resonance strength for the $E_{cm} \approx 480$ keV resonance in the $^{23}$Mg$(p,\gamma)^{24}$Al reaction. Using the thick-target yield equation [24], the resonance strength is given by

$$\omega\gamma = \frac{2Y\epsilon}{\lambda^2}\frac{m_t}{m_b + m_t}, \quad (5)$$

where $\lambda$ is the de Broglie wavelength in the center-of-mass system, $m_b$ and $m_t$ are the respective $^{23}$Mg$^{(5+)}$

and [1]H masses, and $\epsilon$ is the lab-frame stopping power. $Y$ is the recoil yield given by $N_r/(N_b\eta)$, where $N_r$ is the number of detected recoils, $N_b$ is the number of incident beam particles, and $\eta$ is the total recoil detection efficiency given by the product of various individual efficiencies,

$$\eta = f_q \eta_{IC} \eta_{MCP} \tau_{MCP}. \quad (6)$$

The meanings of the various symbols in Equation 6, along with the numerical values of all quantities going into the resonance strength calculation, are presented in Table 2. The separator transmission efficiency is excluded from the efficiency total as it was estimated to be 100% based on the 3.35 mrad cone angle of the $^{24}$Al recoils, which is significantly smaller than the nominal 21 mrad DRAGON acceptance. The number of beam particles given in the table is calculated from the NaI and HPGe decay-station detectors (see Section III.A) but excludes a correction for the data-acquisition live time of 73.5%, for reasons explained at the end of Section III.A. The IC and MCP efficiencies, along with the MCP transmission, were determined as explained in Section III.B. The Al charge state fraction was adopted from Ref. [23], as explained previously in Section III.A.

While the present resonance-strength analysis focuses on the yield of $^{24}$Al obtained in heavy-ion singles, recoil + $\gamma$-ray coincidences were used both to cross-check the identification of the singles $^{24}$Al recoils, as well as to select events for the resonance-energy analysis described in Section III.D. Panel (b) of Figure 9 summarizes the identification of heavy-ion + $\gamma$-ray coincidence recoils. These were identified based on MCP TOF vs. separator TOF, as plotted in the figure. An initial cluster of three events, circled in the figure, was identified with the expected MCP TOF of approximately 54 ns, sitting above a small background from random coincidences. Closer inspection of this cluster (shown in the inset) indicates that the event with separator TOF just below 2420 ns is attributable to a random coincidence between a background $\gamma$-ray and a $^{24}$Al recoil. This conclusion is based on three additional pieces of information. First, the expected separator TOF window for $^{24}$Al recoils is 15 ns, based on GEANT3 simulations of the recoil velocities after the target. The size of this 15 ns window is indicated by the vertical dashed lines in the inset, and even in the extreme case where the other two events sit at the very rightmost edge of the window, the third event would fall outside of the window. Second, the excluded event was found to have a RF$-\gamma$-ray TOF of $-15.3$ ns, relative to the beam crossing the center of the target. This would place the reaction position around 152 mm upstream of the target center. This is far outside the gas region, which extends 61.5(15) mm upstream of the target center. Finally, the energy deposited by this $\gamma$ ray was 2.28 MeV, in a single BGO detector. This energy is not consistent with the expected energy deposition of the primary 1845 keV $\gamma$ ray from the resonance of interest. If the event were genuine, it could only arise from summing of the primary $\gamma$ ray with a secondary 500 keV or 426 keV $\gamma$ ray. Such summing is not likely; for example, GEANT3 simulations of the DRAGON BGO detectors indicate that single-crystal summing between 1845 keV and 500 keV or 426 keV $\gamma$ rays is only expected in around 1.2% of detected events. In contrast, the other two events in the cluster have deposited energies of 1.57 MeV and 1.61 MeV, which are consistent with either photo-absorption or backward-angle Compton scattering by the primary 1845 keV $\gamma$ ray, after accounting for detector resolution.

Based on the arguments given above, only the two events with around 2435 ns separator TOF were tagged as recoil + $\gamma$-ray coincidences and used for further analysis. Each of these events was registered in a BGO detector. The IC anode 1 vs. MCP TOF of these events is shown as the blue filled circles in Figure 9(b) and is consistent with the previously-identified $^{24}$Al locus – further confirming the recoil-singles particle identification. These two events were subsequently analyzed to determine the $^{23}$Mg$(p,\gamma)^{24}$Al resonance energy, as explained in the following subsection.

**D. Resonance Energy**

Using the newly developed resonance timing technique, introduced in Ref. [17], we have extracted an energy for the $E_{cm} \approx 480$ keV $^{23}$Mg$(p,\gamma)^{24}$Al resonance. The resonance timing technique uses the phase of the 11.79 MHz radiofrequency quadrupole (RFQ) signal relative to the detection time of the highest-energy $\gamma$ ray hit to determine the position of the reaction within the 123(2) mm extended gas target. This position is related to the resonance energy due to the continuous energy loss of the beam within the target. In Ref. [17], we characterized the performance of the resonance timing technique using a measurement of the $E_{lab} = 512.1(3)$ keV resonance in the $^{23}$Na$(p,\gamma)^{24}$Mg reaction, with the same $\gamma$-ray detectors used in the present experiment. Using a bootstrap analysis, we showed that the method can be employed with low sample sizes (down to three events), with $\Delta E/E$ uncertainties at the 1% level or better. We also demonstrated that appropriate coverage properties for a $1\sigma$ (68.3%) confidence interval are obtained by quoting the TOF as $T_r \pm \sigma_{T_r}$. Here, $T_r$ is the mean RFQ to $\gamma$-ray TOF, with appropriate corrections for timing offsets. The uncertainty $\sigma_{T_r}$ is given by $\sigma_{T_r} = A s_{T_r}/\sqrt{N}$, with $s_{T_r}$ the sample standard deviation, $N$ the number of coincidences, and $A$ the 84.1% quantile of the Student's $T$ distribution with $N-1$ degrees of freedom.

In the present experiment, we collected only two $^{24}$Al + $\gamma$-ray coincidences, with both $\gamma$-ray events being detected in BGO detectors. Hence, we first extended the bootstrap analysis of Ref. [17] to sample sizes of $N = 2$. For events detected only in the BGO detectors, and including the walk correction explained in Ref. [17], we found that the average confidence interval is given by $\sigma_{T_r} = 2.6$ ns, with a standard deviation of 2.2 ns across all bootstrap trials. The resulting coverage probability was 70.5% – a modest level of over-coverage. This indicates that reliable results for the resonance energy can be extracted by performing the same analysis for the two observed coincidences.

The resonance timing analysis requires that the offsets of the $\gamma$-ray timing signals be synchronized such that $t = 0$ represents the time at which the beam crosses the target center. For the present analysis, the timing offsets determined from the known $E_{lab} = 512.1(3)$ keV resonance in the $^{23}$Na$(p,\gamma)^{24}$Mg reaction, reported in Ref. [17], were used to synchronize the $\gamma$-ray signals. The offsets determined in Ref. [17] utilized a timing walk correction which shifted the mean times observed within 1 MeV energy bins such that they are all synchronized. For the present analysis, this correction was revised to account for the rapid timing shifts observed in the 1-2 MeV bin, which encompasses both detected $\gamma$-ray energies. For the revised walk correction, we fit a plot of time vs. energy, from the $^{23}$Na$(p,\gamma)^{24}$Mg data, with the function

$$f(E) = p_0 - p_1 \exp\left(p_2/\sqrt{E}\right), \qquad (7)$$

which was taken from Ref. [25]. We then evaluated $f(E)$ at the detected $\gamma$-ray energies to obtain the walk correction. We estimated the uncertainty on this correction as 0.33 ns by propagating the uncertainties on the parameter values determined in the fit.

The timing offset also requires an additional correction for the offset between the RFQ signal and the time the beam crosses the target center, which is different between the $^{23}$Na$(p,\gamma)^{24}$Mg experiment and the present one. This offset was obtained by sending attenuated beam directly into a plastic scintillator located downstream of the target, during both the $^{23}$Na$(p,\gamma)^{24}$Mg and $^{23}$Mg$(p,\gamma)^{24}$Al beam times. The resulting scintillator time difference of 10.9(2) ns between the two experiments was used to correct the present TOF values. With this correction, together with the previously explained walk correction, we determined an average TOF for the coincidence events of $-1.075(261)$ ns, indicating a resonance position just upstream of the center of the target. The uncertainty on this quantity is statistical only, obtained from the Student's $T$ distribution as explained in Ref. [17]. Folding in the various estimates of systematic uncertainties results in a TOF of $-1.075(1127)$ ns. The systematic uncertainties included in this calculation are as follows:

1. Uncertainty on the RFQ timing offset measured using the plastic scintillator: 0.2 ns.
2. Uncertainty on the $\gamma$-ray detector timing offsets: 0.6 ns (see Ref. [17]).
3. Uncertainty on the walk correction: 0.33 ns.
4. Uncertainty on the $^{23}$Na$(p,\gamma)^{24}$Mg resonance time: 0.839 ns.

The $-1.075(1127)$ ns TOF translates into a mean $z$-position for the reaction of $-10.5(110)$ mm, using the relativistic equations given in Ref. [17]. The 11.0 mm uncertainty on the position corresponds to 8.96% of the 123(2) mm target effective length. The $z$-position uncertainty was obtained by folding the statistical and systematic uncertainties on the resonance TOF together with the uncertainties on the beam energy, stopping power, target pressure, target temperature, and target effective length (see Section III.A and Table 2 for values of the relevant parameters). Translating the resonance position into energy, again using the relativistic equations in Ref. [17], results in an energy for the key astrophysical $^{23}$Mg$(p,\gamma)^{24}$Al resonance of $478.8(26)$ keV. The various sources of uncertainty impacting the resonance energy and their overall contribution to the error budget are listed in Table 3. The dominant sources of uncertainty are the systematic uncertainties related to the timing offset, as well as the uncertainty on the $^{23}$Mg beam energy.

## IV. DISCUSSION

A summary of published strength and energy values for the $E_{cm} \approx 480$ keV $^{23}$Mg$(p,\gamma)^{24}$Al resonance is given in Table 4. The present strength value of 38(7) meV, extracted from the yield of recoil singles, is the most precise measurement to date. It is in excellent agreement with the previous direct measurement, $38(^{+21}_{-15})$ meV [9], as well as the re-analysis of that work by Wrede *et al.,* $26.6(^{+154}_{-70})$ meV [16]. Our result is also in excellent agreement with the various published shell-model calculations of the resonance strength, when accounting for the approximately 50% uncertainties that are typically associated with shell-model calculations of radiative widths. Our result is in relatively poorer agreement with the 19(9) meV reported by Vyfers *et al.*, although the values are still consistent within $2\sigma$.

The present resonance energy of 478.8(26) keV is in excellent agreement with the 481.0(14) keV

indirect measurement of Lotay *et al.* [6], when the latter is calculated using the AME-20 proton threshold [26]. On the other hand, our result only agrees with Erikson *et al.* at just below the $2\sigma$ level, taking the revised Erikson energy of $485.0(^{+13}_{-18})$ keV. A global view of the various published resonance energies is shown in Figure 10(a). The measurements generally show the expected statistical spread, although the earliest two ($^3$He, $t$) measurements may be considered outliers. We calculated a weighted average value for all measurements by performing a constant-value least-squares fit, using the "soft_l1" loss function provided by the iminuit minimization package [27]. This loss function is quadratic for small deviations between the model and data points ($\ll 1\sigma$), linear for large deviations ($\gg 1\sigma$), and smoothly interpolated in between. The use of this loss function de-emphasized outlier points. We used the MINOS algorithm [28] to determine asymmetric error bars on the fit parameter. The resulting average energy is $481.5(^{+16}_{-13})$ keV, shown by the grey shaded band in the figure. Figure *10*(b) shows the various published resonance strengths, plotting the same values that are listed in Table 4.

Using the RatesMC code [29], we have performed a Monte Carlo calculation of the $^{23}$Mg$(p,\gamma)^{24}$Al stellar rate, which incorporates the results of our measurement and compares it with the recent "ETR25" evaluated rate published by Iliadis *et al.* [30]. The ETR25 compilation evaluates the $^{23}$Mg$(p,\gamma)^{24}$Al rate using the same RatesMC code as the present calculations. For the key $E_{cm} \approx 480$ keV resonance, the ETR25 authors adopt an energy of 482.13(142) keV. They treat the strength of this resonance as "unmeasured," meaning that the resonance strength is calculated indirectly using estimated proton and radiative partial widths and the equation $\omega\gamma = (2J+1)\Gamma_p\Gamma_\gamma/[8(\Gamma_p+\Gamma_\gamma)]$, with $J=3$ the known resonance spin. The partial widths were taken from mirror-state information, with adopted values of $\Gamma_p = 220$ meV and $\Gamma_\gamma = 45.6$ meV. The uncertainty on both widths was taken to be a multiplicative factor of 2. Other $^{23}$Mg$(p,\gamma)^{24}$Al resonances were treated similarly in the ETR25 calculations. For our present rate calculations, we use the ETR25 RatesMC inputs for direct capture and all resonances except the $E_{cm} \approx 480$ keV resonance. For the $E_{cm} \approx 480$ keV resonance, we use the presently determined strength of $38(7)$ meV and the weighted-average energy of $481.5(^{+16}_{-13})$ keV presented earlier in this section (RatesMC does not have an option to handle asymmetric error bars, so we used a symmetric uncertainty of 1.6 keV as the input).

The results of our rate calculations are presented in Table 5, which lists the "low" (16th percentile), "median" (50th percentile), and "high" (84th percentile) rates as a function of temperature in GK ($T_9$). The table also lists the factor uncertainty (f.u.) as a function of temperature. This quantity is derived from a lognormal approximation of the Monte Carlo samples: $\text{f.u.} = e^{\sigma}$, where $\sigma$ is the lognormal width parameter. We also include a machine-readable table of the full RatesMC output as Supplemental Material. In addition to the low, median, and high rates as a function of $T_9$, the supplemental table also lists, as a function of $T_9$, the "$2\sigma$" (2nd and 98th percentile) low and high rates; the mean and "classical" rates; and the location and width parameters, along with the Anderson Darling statistic, of a lognormal fit to the Monte Carlo samples.

The results of our rate calculations are also summarized in Figure 11. The figure includes a plot of the total median rate in panel (a), as well as the individual median direct-capture and resonance contributions. The figure also shows, in panel (b), the fractional contributions of direct capture and the various resonances to the total rate, including uncertainty bands corresponding to the $1\sigma$ (16th/84th percentile) low/high rates. We note that while the $E_{cm} \approx 480$ keV resonance dominates across nearly the entire nova temperature range of 0.2-0.4 GK, it has only an approximate 50% fractional contribution near the lower end of this range, with the remainder of the rate at these temperatures dominated by direct capture. Panel (c) of the figure shows the $1\sigma$ (16th/84th percentile) and $2\sigma$ (2nd/98th percentile) uncertainties on the total rate, expressed as the ratio of the corresponding low or high rate to the median. In the nova temperature range of 0.2-0.4 GK, the $1\sigma$ reaction rate uncertainty corresponds to a total up/down factor of approximately 1.5 or less, meaning that the reaction is no longer a likely dominant source of uncertainty in classical nova nucleosynthesis. Finally, Panel (d) displays the ratio of the present rate to the ETR25 rate. The solid line displays the ratio of median rates (present/ETR25), while the dark and light shaded regions represent the ratio of the present median rate to the $1\sigma$ (dark) and $2\sigma$ (light) low and high rates. Comparing with Panel (c), one can clearly see that the present uncertainty bands are substantially smaller than ETR25. The two rate calculations also agree within the $1\sigma$ uncertainty band, which is consistent with the good agreement between the ETR25 mirror-state inputs for the $E_{cm} \approx 480$ keV resonance and those determined in the present work.

## V. SUMMARY

We have performed a new inverse-kinematics direct measurement of the key astrophysical $E_{cm} \approx 480$ keV resonance in the $^{23}$Mg$(p,\gamma)^{24}$Al reaction. The measurement used a radioactive $^{23}$Mg beam with minimal $^{23}$Na contamination, at an energy of $E/A = 507.6(10)$ keV, which centers the resonance in the extended $H_2$ reaction target. The measurement is also the first application of the resonance timing technique in a radioactive beam experiment.

From the singles yield of $^{24}$Al recoils, we determined a resonance strength of 38(7) meV. This is in good agreement with the earlier direct measurement by Erikson *et al.*, as well as shell model calculations. We also measured the energy of this resonance to be $E_{cm} = 478.8(26)$ keV, using the resonance timing method. This is in excellent agreement with the level energy reported by Lotay *et al.*, combined with the proton threshold from the AME-20 compilation. A global fit of all published resonance energies (including the present work), using a cost function that minimizes the role of outliers, results in a recommended energy of $E_{cm} = 481.5(^{+16}_{-13})$ keV.

Incorporating our results into Monte Carlo calculations of the stellar $^{23}\mathrm{Mg}(p,\gamma)^{24}\mathrm{Al}$ rate indicates that the reaction is well constrained, to a factor of approximately 1.5 or better across the relevant temperature range for novae. The key remaining uncertainty in this reaction is likely to be the direct-capture contribution, which may be significant near 0.2 GK. With future increases in $^{23}$Mg beam intensities, a direct measurement of direct capture may become feasible down to energies as low as $E_{cm} \approx 500$ keV. Such a measurement would require a beam intensity of around $10^{10}$ s$^{-1}$, which is approximately an order of magnitude larger than current best yields at the TRIUMF-ISAC facility. With the anticipated increases in beam-development time expected after the ARIEL facility comes online at TRIUMF, such an experiment may soon become feasible.

## ACKNOWLEDGMENTS

The authors are grateful to the TRIUMF beam-delivery group for their efforts throughout the experiment. TRIUMF's core operations are supported by a federal contribution through the National Research Council of Canada, and the Government of British Columbia provides building capital funds. GC, ACE, ED, and JO acknowledge the support of the Natural Sciences and Engineering Research Council of Canada (NSERC), awards SAPIN-2020-00052 and RGPAS-2020-00001. DRAGON is funded through NSERC award SAPPJ-2023-00039. PHR, ZP and SP were supported by the UK STFC UK Nuclear Data Network and the UK Science and Technology Facilities Council (STFC) via Grant Numbers ST/V001108/1 and ST/Y000358/1. PHR, RS and SMC acknowledge support from the UK National Measurements System Programmes Unit of the UK's Department for Science, Innovation and Technology (DSIT). UG acknowledges that this material is based upon work supported by the U.S. DOE, Office of Science, Office of Nuclear Physics under contract DE-FG02-93ER40789.

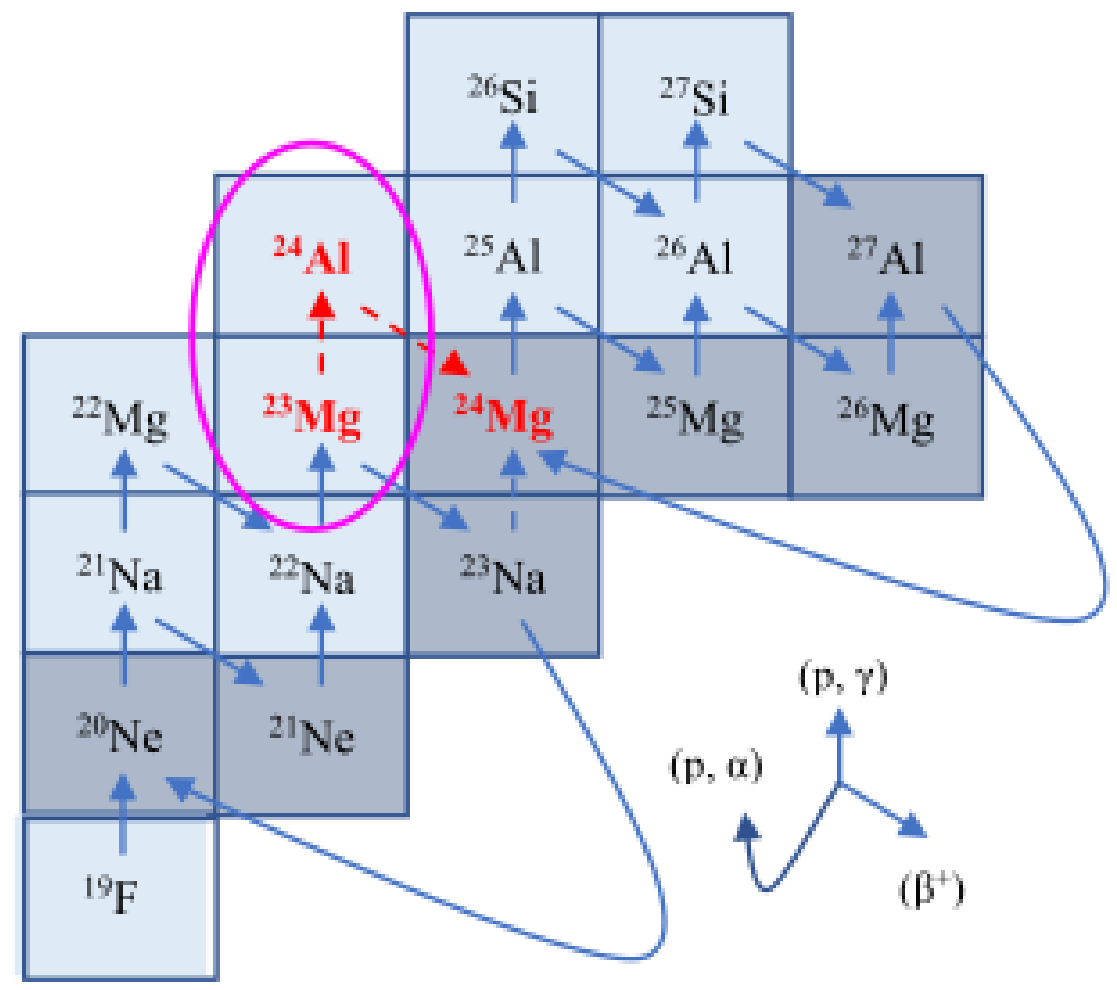

Figure 1: Selected portion of a typical nova reaction network, showing the Ne-Na and Mg-Al cycles. The $^{23}$Mg(p,$\gamma$)$^{24}$Al reaction is circled in magenta. The nuclides shown with darker shading are stable, while those with the lighter shading are $\beta$-unstable.

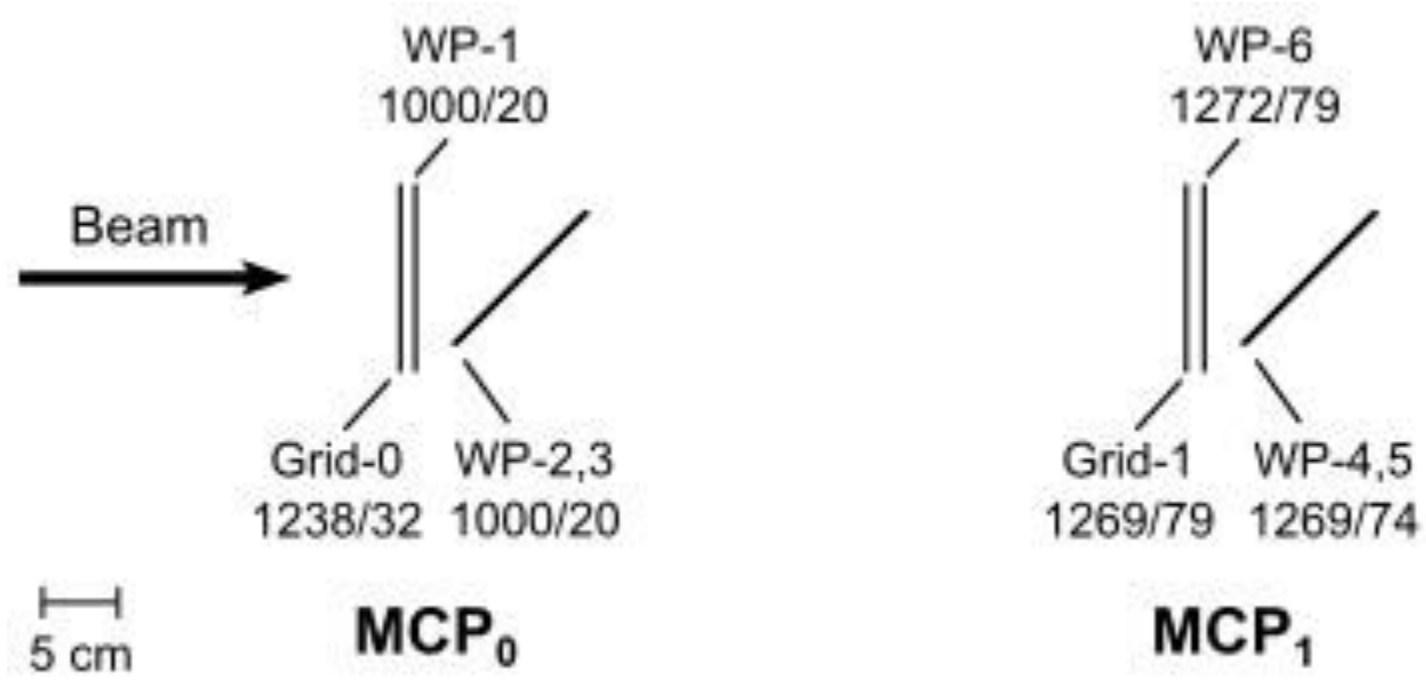

Figure 2: Schematic of the MCP elements traversed by the beam. The numbers below each grid or wire plane (WP) element represent the wire pitch/thickness in $\mu$m (for example, WP-1 has a pitch of 1000 $\mu$m and thickness 20 $\mu$m). The tilted WP elements are all at a 45$^o$ angle relative to the beam axis. The wires on both grids, as well as WPs 4, 5, and 6, are two-dimensional meshes, while only vertical wires are present on WPs 1, 2, and 3.

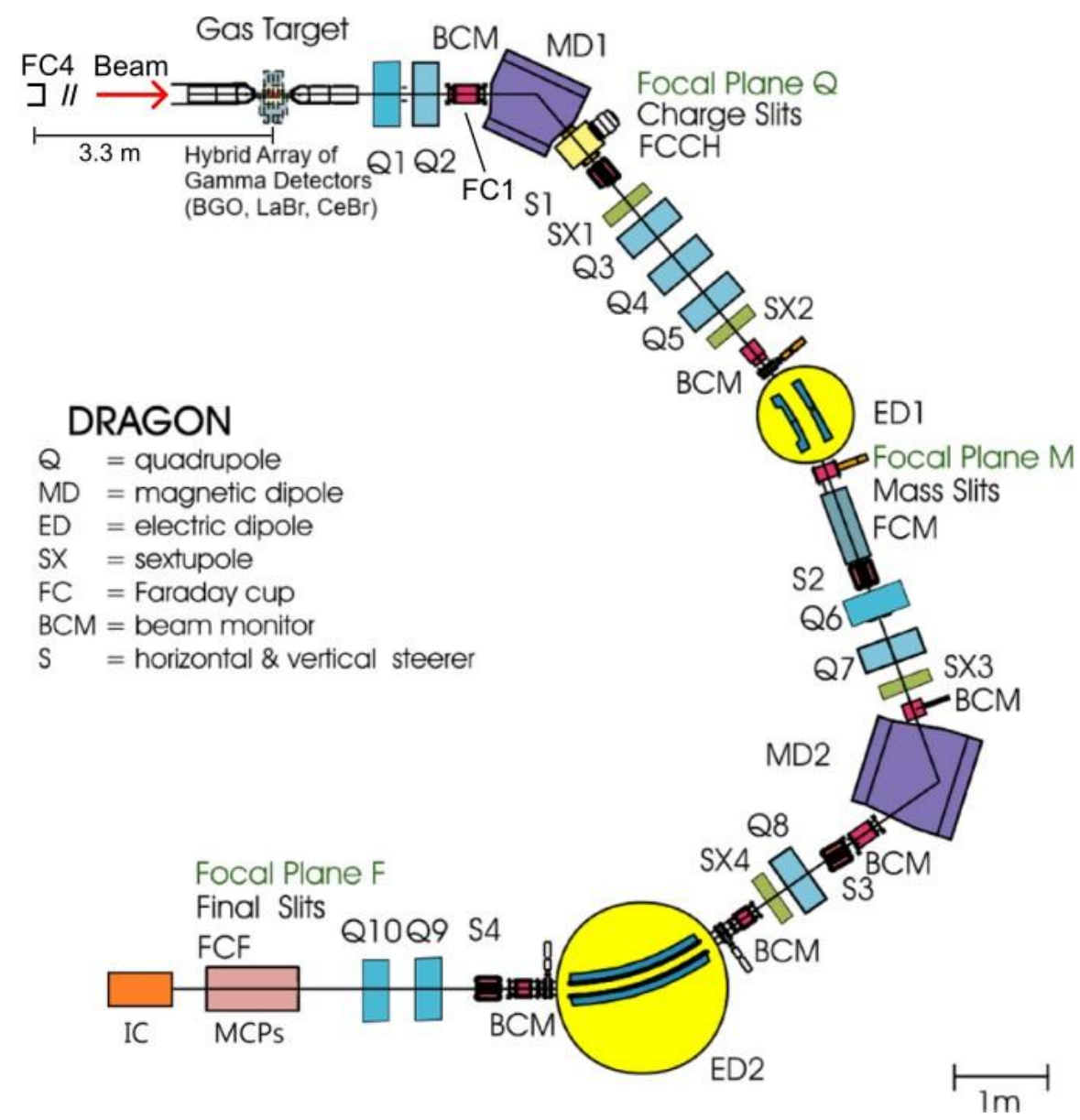

Figure 3: Schematic of the DRAGON system, showing the placement of various optical elements, as well as the $\gamma$-ray and charged-particle detectors, as used in the present work.

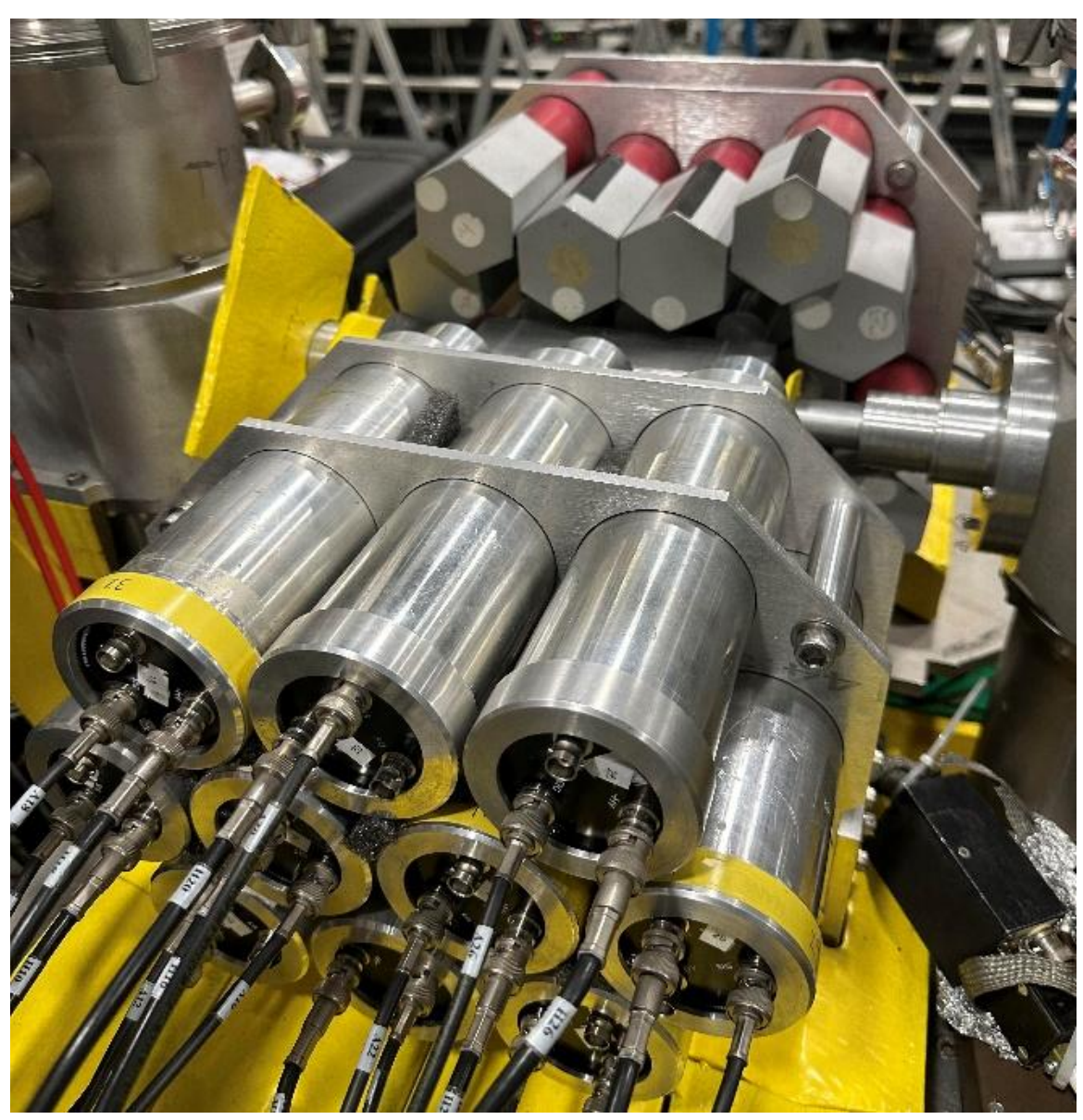

Figure 4: Photograph of the hybrid $\gamma$-ray detector array surrounding the DRAGON gas target.

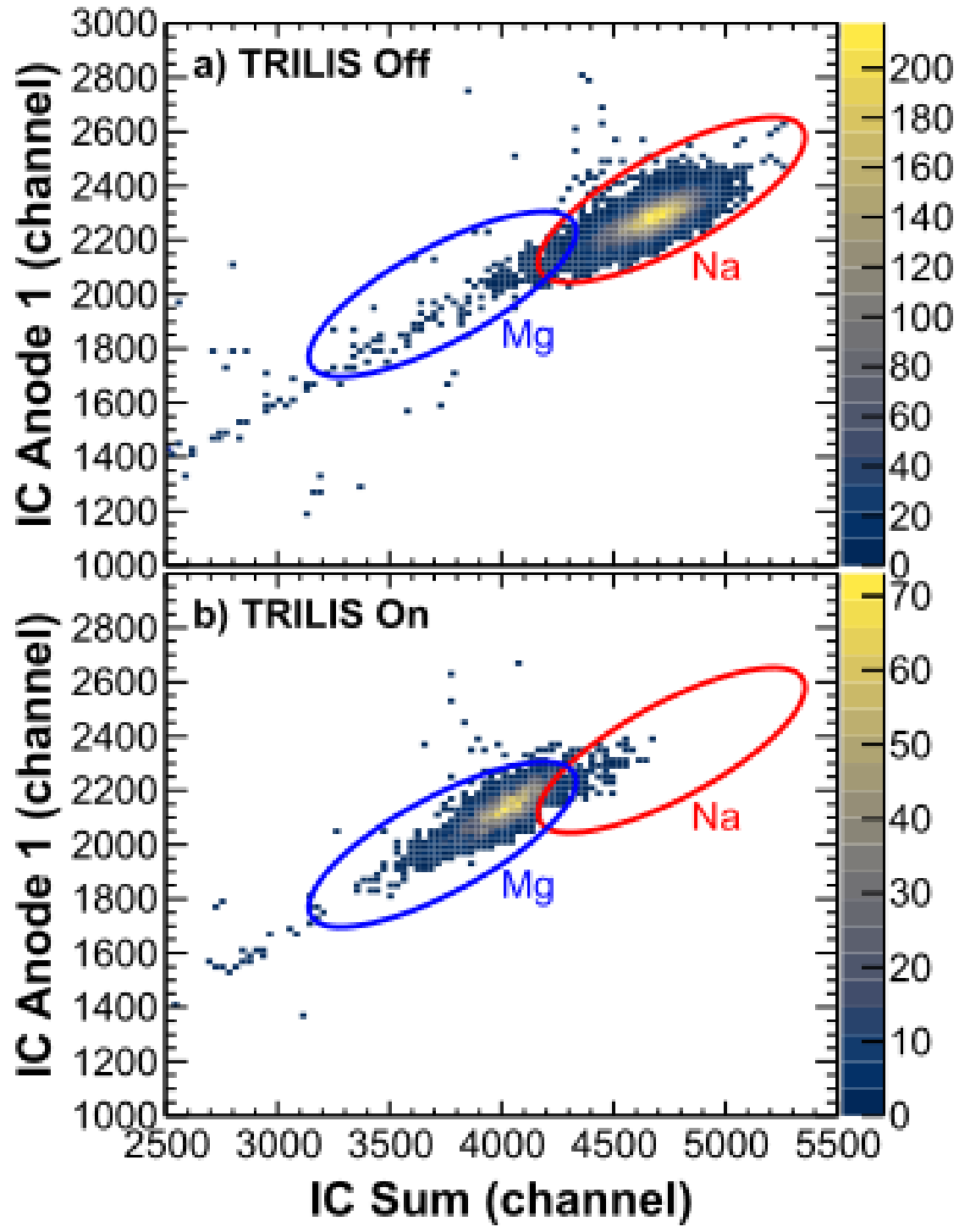

Figure 5: IC energy loss signals for attenuated beam sent directly into the detector. Panel a) shows the IC spectrum with TRILIS lasers off, and panel b) shows the spectrum with TRILIS lasers on.

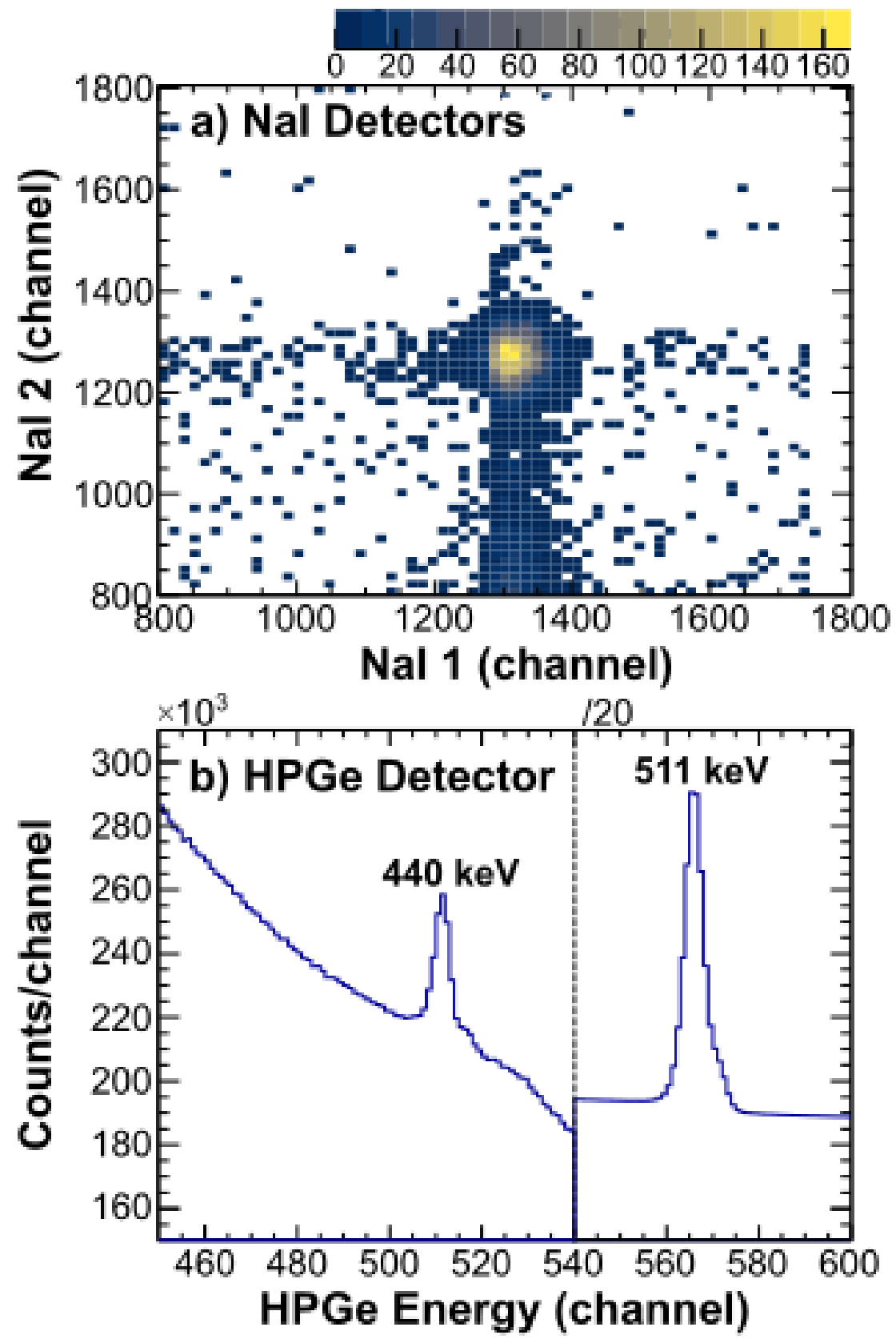

Figure 6: a) Coincident energy spectrum (in channel number) from the mass-slit NaI detectors. The bright locus between channels 1200-1400 on both axes consists of 511 keV $\gamma$-$\gamma$ coincidences. b) Energy spectrum (in channel number) from the mass-slit HPGe detector. For display purposes, counts to the right of the dashed line are scaled down by a factor 20, and the baseline is shifted up by approximately $200 \times 10^3$ counts/channel. The location of the 440 keV and 511 keV lines is indicated on the figure.

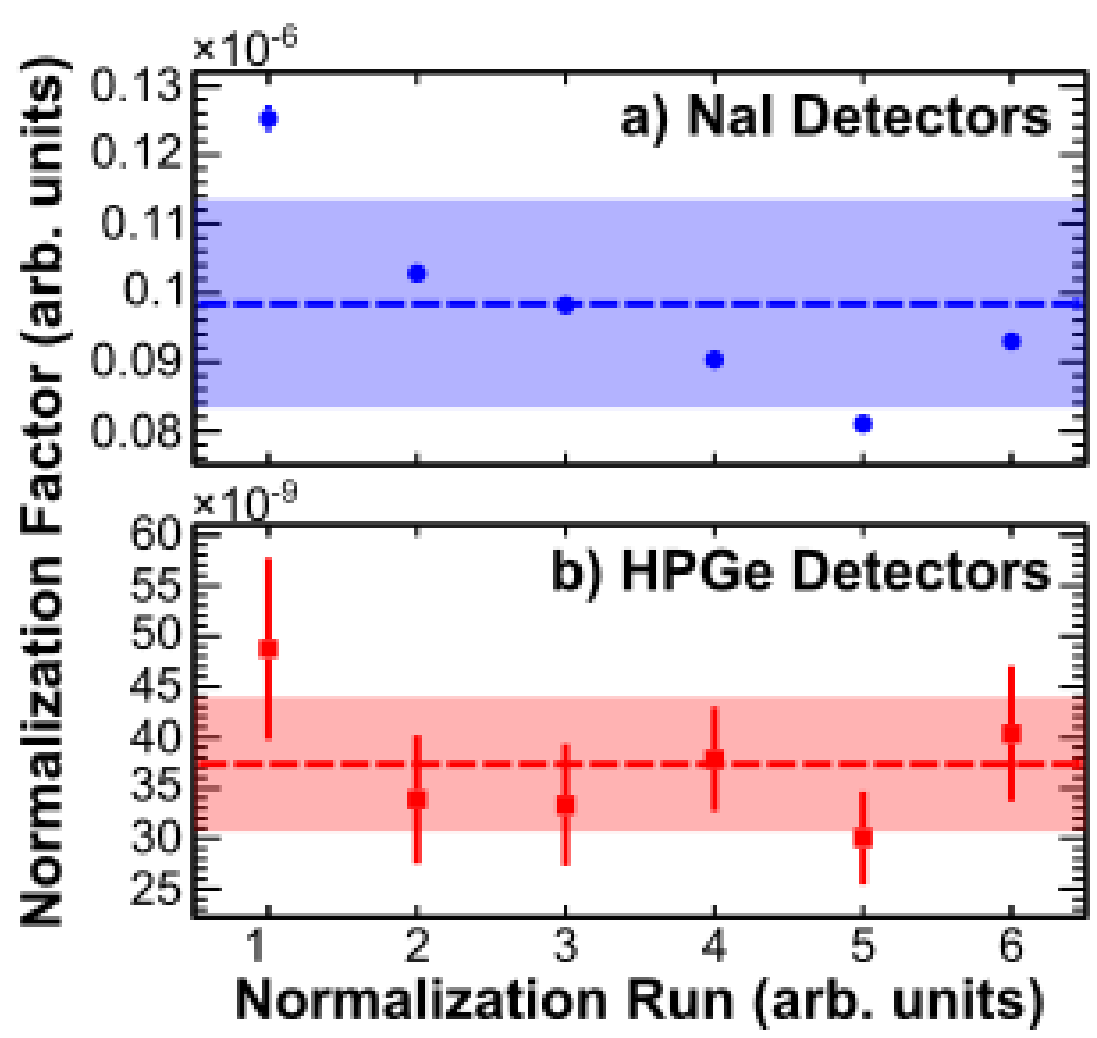

Figure 7: Normalization factor $\rho$ (see text) determined for a) the coincident NaI system and b) the HPGe 440 keV $\gamma$ ray. In each panel, the dashed lines and shaded regions represent the mean and standard deviation of the various factors, respectively. Error bars are statistical only.

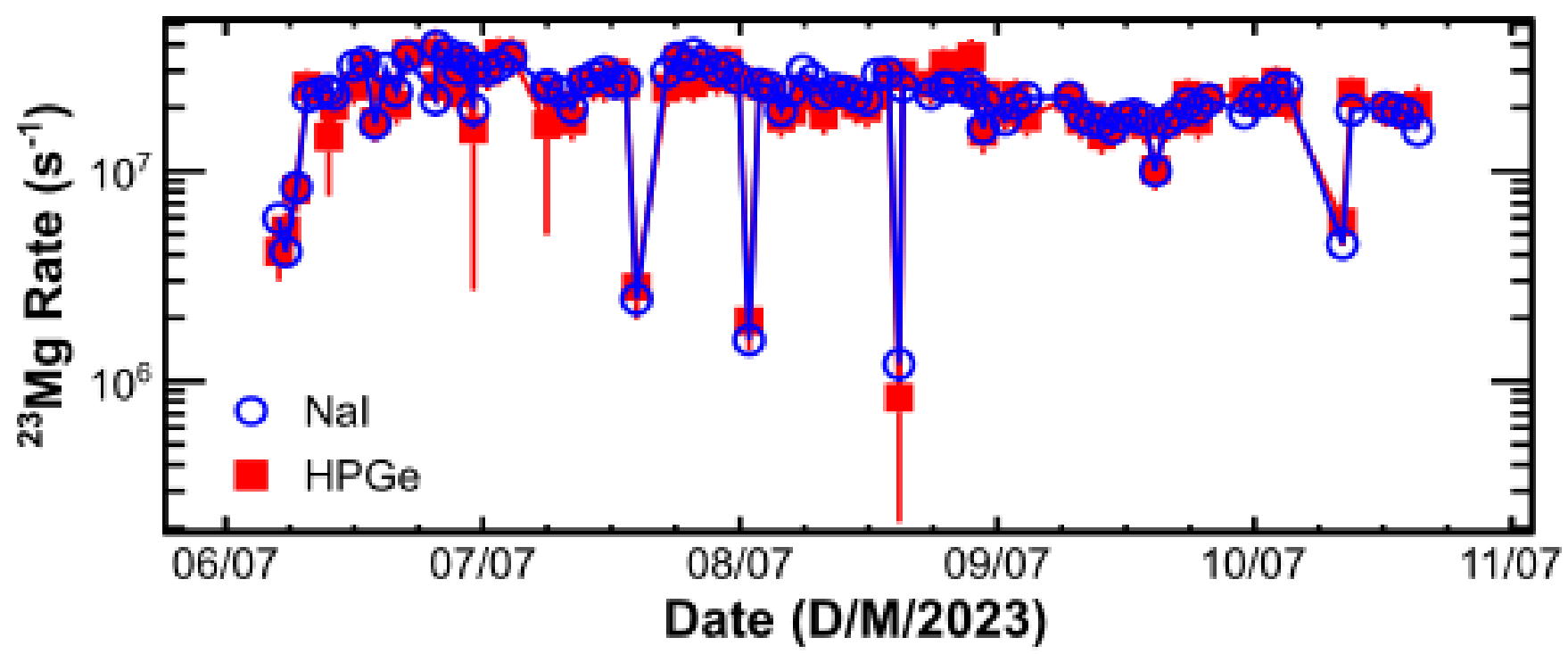

Figure 8: Incident $^{23}$Mg beam rate determined using the coincident NaI detectors (blue open circles) and the HPGe detector (red filled squares).

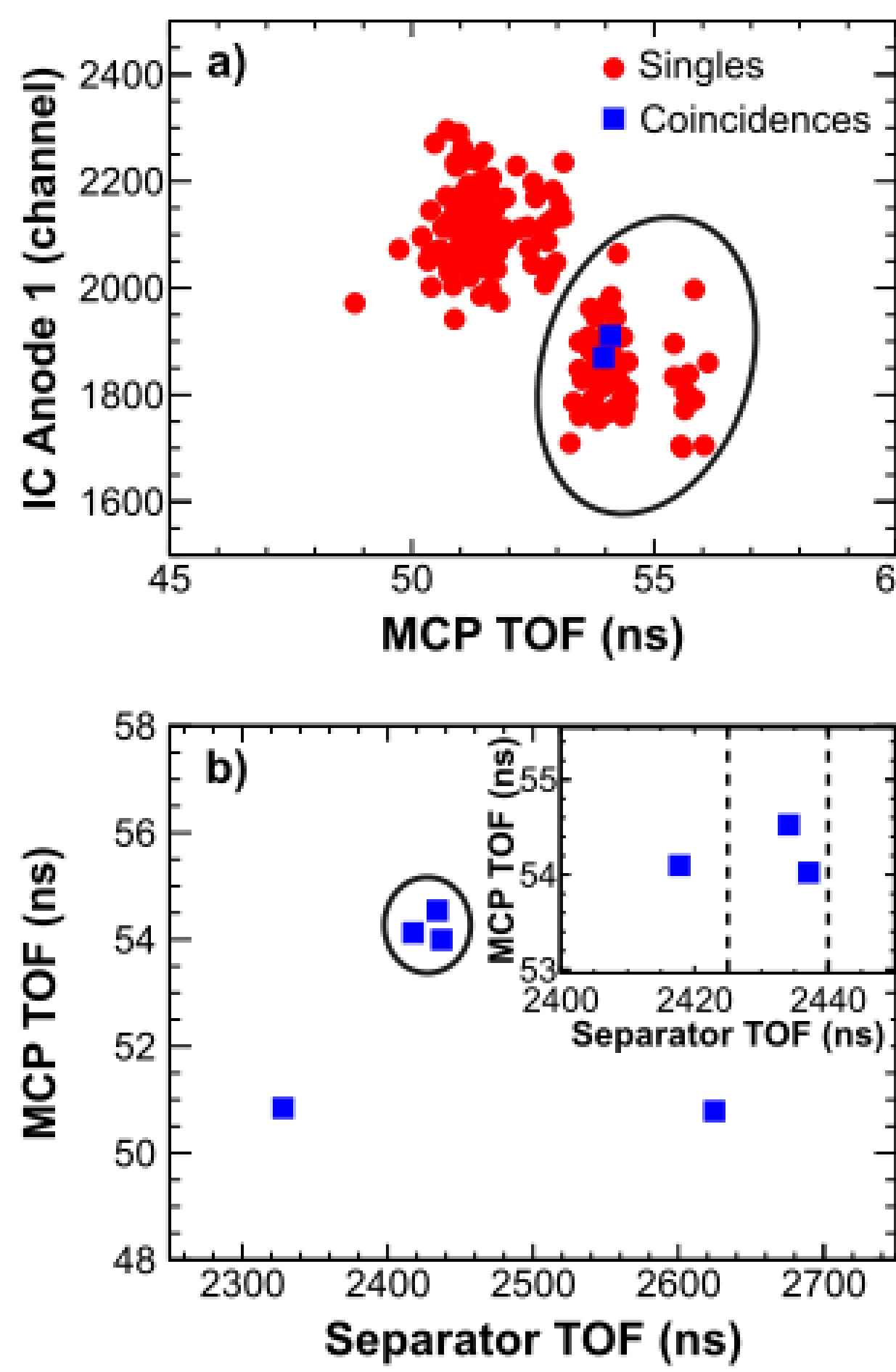

Figure 9: Recoil identification for both heavy-ion + $\gamma$-ray coincidences and heavy-ion singles. Panel a) shows the IC energy deposition in Anode 1 vs. MCP TOF, for both recoil singles (red circles) and for events passing all coincidence cuts (blue squares). Panel b) shows separator vs. MCP TOF, for coincidences. The inset displays the same plot, zoomed into the cluster of three events circled in the main panel. The dashed vertical lines in the inset represent the expected 15 ns width of separator TOF of the recoils.

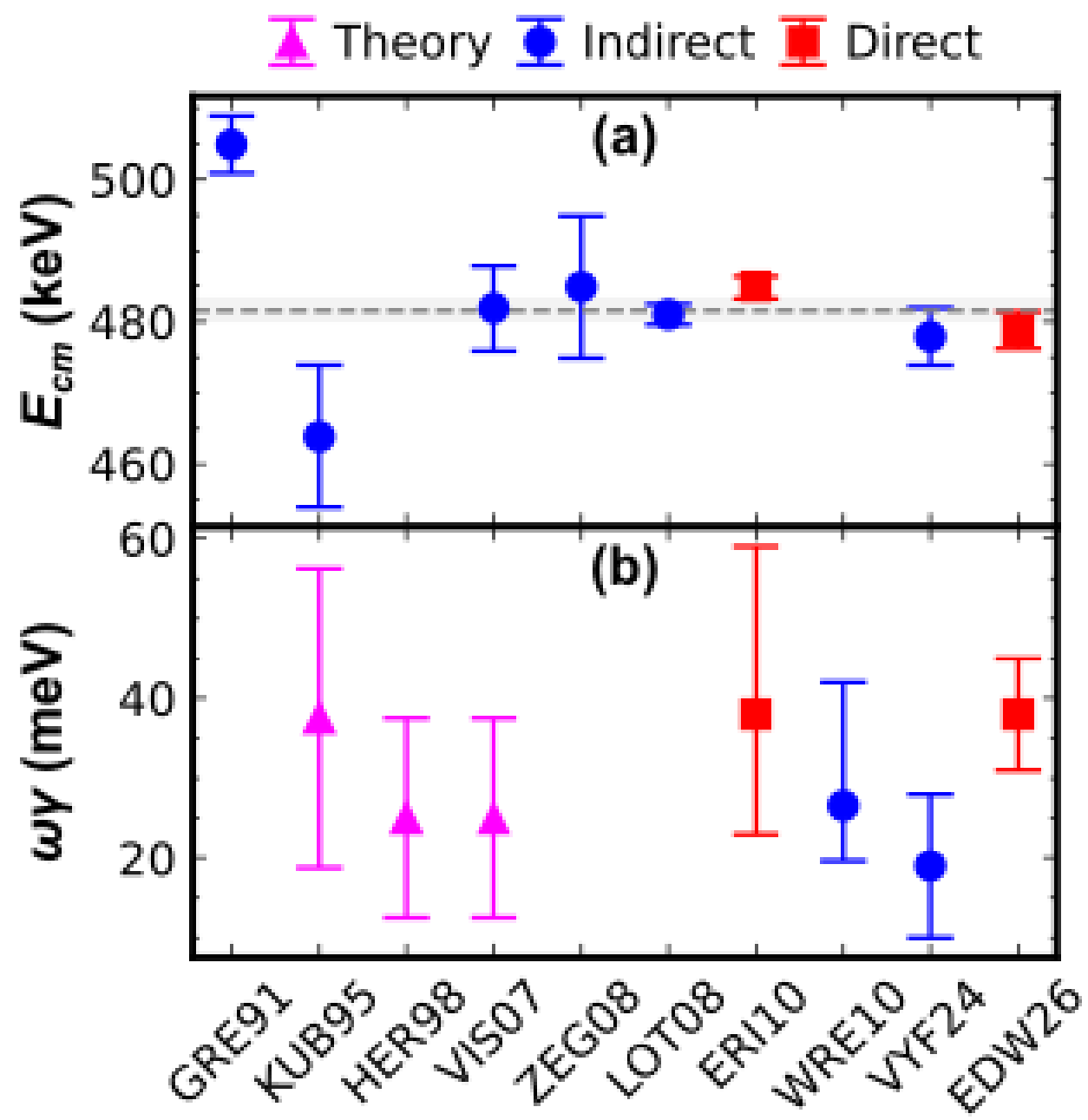

Figure 10: Summary of published resonance strength and energy values for the $E_{cm} \approx 480$ keV resonance in $^{23}$Mg$(p,\gamma)^{24}$Al. The plotted values are identical to those given in Table 4, with the panel (a) displaying resonance energies and panel (b) displaying strengths. The fuchsia triangles, labeled "Theory" represent published strengths that are primarily based on theoretical calculations. The error bars for these points are estimated at 50% of the central value. Filled blue circles represent energy or strength values primarily based on indirect measurements, while the filled red squared represent direct measurements. In the panel (a) the grey dashed line and shaded band show our recommended resonance energy of $481.5(^{+16}_{-13})$ keV, described in the main text. The sources listed in the figure are as follows: GRE91 [2], KUB95 [3], HER98 [11], VIS07 [4], ZEG08 [5], VYF24 [7], LOT08 [6], ERI10 [9], WRE10 [16], VFY24 [7], and EDW26 [present work].

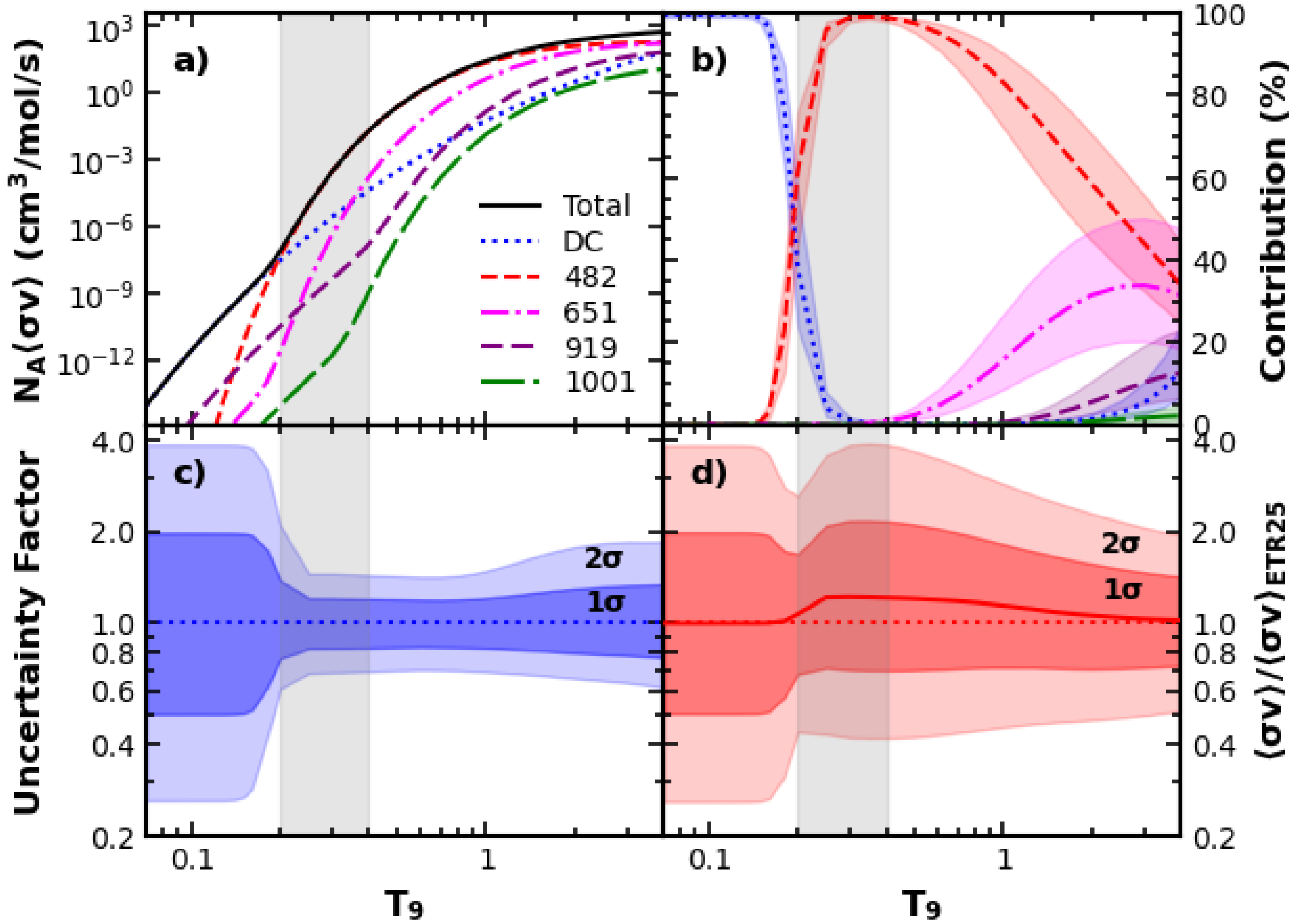

Figure 11: Summary of the reaction rate calculations performed using the present strength and our recommended weighted-average energy for the $E_{cm} \approx 480$ keV resonance. Panel (a) shows the total reaction rate, as well as the individual contribution of various resonances and direct capture. The various curves represent median rate values from the RatesMC calculation, with the resonance energies indicated in the legend. Panel (b) shows the fractional contribution of the various resonances and direct capture. The curves represent the median contribution, and the shaded regions indicate $1\sigma$ uncertainties. Panel (c) shows the $1\sigma$ (dark blue shading) and $2\sigma$ (light blue shading) uncertainties on the total rate, expressed as the ratio of the corresponding upper or lower rate to the median rate. Panel (d) shows the ratio of the present median rate to the ETR25 rate [30]. The solid red line represents the radio of median rates, while dark and light shaded regions represent the ratios of the present median rate to the respective $1\sigma$ and $2\sigma$ low and high ETR25 rates. The vertical grey shaded regions on each plot represent the relevant nova temperature range of 0.2–0.4 GK.

Table 1: Summary of Mg:Na ratios obtained using attenuated beam sent directly into the IC detector.

| Run Number | Mg:Na Ratio |
|---|---|
| 11505 | 46.0 |
| 11527 | 6.61 |
| 11558 | 94.4 |
| 11588 | 17.1 |
| 11625 | 19.4 |
| 11642 | 16.0 |

Table 2: Summary of the quantities going into the yield and resonance strength calculations. As mentioned in the footnotes, the live time correction is excluded from both the number of beam particles and the recoil detection efficiency, as these cancel in the yield ratio. The relative uncertainties listed in the rightmost column are calculated prior to rounding off the nominal and uncertainty values of each parameter.

| Parameter | Description | Value | Relative Uncertainty |
|---|---|---|---|
| $N_r$ | Number of recoils. | 63(8) | 12.6% |
| $N_b$ | Number of incident $^{23}$Mg.[1] | $4.7(6) \times 10^{12}$ | 11.7% |
| $m_p$ | $^{1}$H mass.[2] | 1.00728 u | — |
| $m_b$ | $^{23}$Mg$^{(5+)}$ mass.[2] | 22.9914 u | — |
| $\lambda$ | de Broglie wavelength (CM frame). | $4.22063 \times 10^{-12}$ cm | — |
| $\epsilon$ | Stopping power (lab frame).[3] | $105.7(11) \times 10^{-15}$ eV cm$^2$ | 1.0% |
| $\eta_{MCP}$ | MCP detection efficiency.[4] | 97.4(4)% | 0.41% |
| $\eta_{IC}$ | IC detection efficiency.[4,5] | 96.0(15)% | 1.6% |
| $\tau_{MCP}$ | MCP transmission.[4] | 50.2(11)% | 2.2% |
| $\tau_{wire1}$ | MCP1 wire-plane transmission.[4] | 72.3(11)% | 1.5% |
| $f_q$ | $^{24}$Al $6^+$ charge state fraction.[6] | 37.4(23)% | 6.0% |
| $\eta$ | Total detection efficiency.[7] | 17.6(12)% | 6.6% |
| $Y$ | $^{23}$Mg(p,$\gamma$)$^{24}$Al yield. | $7.6(14) \times 10^{-11}$ | 18.5% |
| $\omega\gamma$ | $^{23}$Mg(p,$\gamma$)$^{24}$Al resonance strength. | 38(7) meV | 18.5% |

[1]Excludes 73.5% live time correction and implicitly includes gas-target transmission factor.
[2]Taken from Ref. [26].
[3]Weighted average of present measurements and those of Ref. [23].
[4]See Section III.B.
[5]The error bar includes the uncertainty on the MCP1 wire plane transmission.
[6]Taken from Ref. [23], corrected for the present beam energy.
[7]Excluding data-acquisition live time.

Table 3: Contributions of various sources of uncertainty to the overall uncertainty budget on the $^{23}Mg(p,\gamma)^{24}Al$ resonance energy. The left column gives a description of each source of uncertainty, and the right column denotes each source's impact on the resonance energy uncertainty.

| **Uncertainty Source** | **Contribution (keV)** |
|---|---|
| Statistical uncertainty. | 0.56 |
| RFQ timing offset. | 0.36 |
| $\gamma$-ray timing offset. | 1.3 |
| Timing walk correction. | 0.71 |
| $^{23}Na(p,\gamma)^{24}Mg$ resonance time. | 1.8 |
| Beam energy. | 0.97 |
| $^{23}Mg + H_2$ stopping power. | 0.12 |
| Target pressure. | 0.21 |
| Target temperature. | 0.04 |
| Target effective length. | 0.22 |
| Total | 2.6 |

Table 4: Summary of published and derived values of the resonance strength and energy of the $E_{cm} \approx 480$ keV resonance in $^{23}Mg(p,\gamma)^{24}Al$.

| **Source** | **Method** | **$E_x$ (keV)** | **$E_r$ (keV)** | **$\Gamma_p$ (meV)** | **$\Gamma_\gamma$ (meV)** | **$\omega\gamma$ (meV)** |
|---|---|---|---|---|---|---|
| Present | $^{23}Mg(p,\gamma)^{24}Al$ | 2342.9(26)[1] | 478.8(26) | — | — | 38(7) |
| Erikson et al. [9] | $^{23}Mg(p,\gamma)^{24}Al$ | $2349.1(^{+13}_{-18})$[1] | $485.0(^{+13}_{-18})$[2] | — | — | $38(^{+21}_{-15})$ |
| Greenfield et al. [2] | $^{24}Mg(^3He,t)^{24}Al$ | 2369(4) | 505(4)[1] | — | — | — |
| Kubono et al. [3] | $^{24}Mg(^3He,t)^{24}Al$ | 2328(10) | 464(10)[1] | 489[3] | 47[4] | 37.5[5,6] |
| Visser et al. [4] | $^{24}Mg(^3He,t)^{24}Al$ | 2346(6) | 482(6)[1] | 173[7] | 33[7] | 25[5] |
| Zegers et al. [5] | $^{24}Mg(^3He,t)^{24}Al$ | 2349(10) | 485(10)[1] | — | — | — |
| Vyfers et al. [7] | $^{24}Mg(^3He,t)^{24}Al$ | 2342(4) | 478(4)[1] | $>0.64\Gamma_{tot}$ | 24[8] | 19(9) |
| Lotay et al. [6] | $^{10}B(^{16}O, 2n\gamma)^{24}Al$ | 2345.1(14) | 481.0(14)[1] | — | — | — |
| Herndl et al. [11] | Shell Model | — | — | 175 | 33 | 25[5] |
| Wrede et al. [16] | Reanalysis | — | — | — | — | $26.6(^{+154}_{-70})$[9] |

[1]Calculated with $S_p = 1864.11(23)$ keV, from the 2020 AME [26].
[2]Calculated using the revised DRAGON magnetic constant from Ref. [15].
[3]Calculated assuming $L = 2$ and $\theta_p^2 = 0.1$.
[4]Taken from the reported $^{24}Na$ mirror-state lifetime in Ref. [31], without energy scaling.
[5]Calculated from the authors' $\Gamma_p$ and $\Gamma_\gamma$ values, using the equation $\omega\gamma = (2J+1)\Gamma_p\Gamma_\gamma/[8(\Gamma_p+\Gamma_\gamma)]$, with $J = 3$.
[6]Table 2 of Ref. [3] lists $\omega\gamma = 26.8$ meV for this resonance. However, the $J = 3$ spin and partial widths listed elsewhere in same table result in $\omega\gamma = 37.5$ meV, while 26.8 meV corresponds to a strength calculated using $J = 2$. We therefore adopt our calculated strength of 37.5 meV for the present tabulation.
[7]Taken from the shell-model calculation of Herndl *et al.* [11], scaled to the resonance energy reported in Ref. [4].
[8]Adopted from the authors' shell-model calculations using the Ormand & Brown Hamiltonian.
[9]Extracted from folding the likelihood surface in Ref. [9] with a resonance energy of 482.1(20) keV.

Table 5: Monte Carlo reaction rates calculated using the present $E_{cm} \approx 480$ keV resonance strength and recommended energy. The quantities in the table are the same as those given in ETR25 [30] and are described in the main text. All rates are presented as $N_A\langle\sigma v\rangle$, in units of cm$^3$ mol$^{-1}$ s$^{-1}$.

| $T_9$ | Low Rate | Median Rate | High Rate | f.u. | $T_9$ | Low Rate | Median Rate | High Rate | f.u. |
|---|---|---|---|---|---|---|---|---|---|
| 0.001 | $3.580 \times 10^{-86}$ | $7.148 \times 10^{-86}$ | $1.418 \times 10^{-85}$ | 1.998 | 0.140 | $1.804 \times 10^{-10}$ | $3.578 \times 10^{-10}$ | $7.076 \times 10^{-10}$ | 1.987 |
| 0.002 | $1.173 \times 10^{-66}$ | $2.342 \times 10^{-66}$ | $4.648 \times 10^{-66}$ | 1.998 | 0.150 | $4.556 \times 10^{-10}$ | $8.980 \times 10^{-10}$ | $1.771 \times 10^{-09}$ | 1.976 |
| 0.003 | $3.475 \times 10^{-57}$ | $6.939 \times 10^{-57}$ | $1.377 \times 10^{-56}$ | 1.998 | 0.160 | $1.099 \times 10^{-09}$ | $2.118 \times 10^{-09}$ | $4.130 \times 10^{-09}$ | 1.939 |
| 0.004 | $3.285 \times 10^{-51}$ | $6.558 \times 10^{-51}$ | $1.301 \times 10^{-50}$ | 1.998 | 0.180 | $7.165 \times 10^{-09}$ | $1.168 \times 10^{-08}$ | $2.046 \times 10^{-08}$ | 1.689 |
| 0.005 | $5.849 \times 10^{-47}$ | $1.168 \times 10^{-46}$ | $2.317 \times 10^{-46}$ | 1.998 | 0.200 | $6.452 \times 10^{-08}$ | $8.412 \times 10^{-08}$ | $1.164 \times 10^{-07}$ | 1.367 |
| 0.006 | $1.022 \times 10^{-43}$ | $2.040 \times 10^{-43}$ | $4.048 \times 10^{-43}$ | 1.998 | 0.250 | $8.324 \times 10^{-06}$ | $1.006 \times 10^{-05}$ | $1.218 \times 10^{-05}$ | 1.211 |
| 0.007 | $4.006 \times 10^{-41}$ | $7.956 \times 10^{-41}$ | $1.575 \times 10^{-40}$ | 1.990 | 0.300 | $2.528 \times 10^{-04}$ | $3.051 \times 10^{-04}$ | $3.690 \times 10^{-04}$ | 1.210 |
| 0.008 | $5.477 \times 10^{-39}$ | $1.088 \times 10^{-38}$ | $2.154 \times 10^{-38}$ | 1.990 | 0.350 | $2.865 \times 10^{-03}$ | $3.453 \times 10^{-03}$ | $4.164 \times 10^{-03}$ | 1.207 |
| 0.009 | $3.497 \times 10^{-37}$ | $6.945 \times 10^{-37}$ | $1.375 \times 10^{-36}$ | 1.990 | 0.400 | $1.735 \times 10^{-02}$ | $2.087 \times 10^{-02}$ | $2.510 \times 10^{-02}$ | 1.205 |
| 0.010 | $1.254 \times 10^{-35}$ | $2.490 \times 10^{-35}$ | $4.932 \times 10^{-35}$ | 1.990 | 0.450 | $6.933 \times 10^{-02}$ | $8.319 \times 10^{-02}$ | $9.984 \times 10^{-02}$ | 1.202 |
| 0.011 | $2.867 \times 10^{-34}$ | $5.693 \times 10^{-34}$ | $1.127 \times 10^{-33}$ | 1.990 | 0.500 | $2.074 \times 10^{-01}$ | $2.484 \times 10^{-01}$ | $2.976 \times 10^{-01}$ | 1.200 |
| 0.012 | $4.573 \times 10^{-33}$ | $9.081 \times 10^{-33}$ | $1.798 \times 10^{-32}$ | 1.990 | 0.600 | $1.048 \times 10^{+00}$ | $1.252 \times 10^{+00}$ | $1.495 \times 10^{+00}$ | 1.196 |
| 0.013 | $5.440 \times 10^{-32}$ | $1.080 \times 10^{-31}$ | $2.139 \times 10^{-31}$ | 1.990 | 0.700 | $3.254 \times 10^{+00}$ | $3.894 \times 10^{+00}$ | $4.644 \times 10^{+00}$ | 1.196 |
| 0.014 | $5.075 \times 10^{-31}$ | $1.008 \times 10^{-30}$ | $1.996 \times 10^{-30}$ | 1.990 | 0.800 | $7.496 \times 10^{+00}$ | $8.981 \times 10^{+00}$ | $1.074 \times 10^{+01}$ | 1.199 |
| 0.015 | $3.859 \times 10^{-30}$ | $7.664 \times 10^{-30}$ | $1.518 \times 10^{-29}$ | 1.990 | 0.900 | $1.415 \times 10^{+01}$ | $1.700 \times 10^{+01}$ | $2.045 \times 10^{+01}$ | 1.206 |
| 0.016 | $2.467 \times 10^{-29}$ | $4.899 \times 10^{-29}$ | $9.701 \times 10^{-29}$ | 1.990 | 1.000 | $2.330 \times 10^{+01}$ | $2.807 \times 10^{+01}$ | $3.403 \times 10^{+01}$ | 1.214 |
| 0.018 | $6.567 \times 10^{-28}$ | $1.304 \times 10^{-27}$ | $2.582 \times 10^{-27}$ | 1.990 | 1.250 | $5.552 \times 10^{+01}$ | $6.762 \times 10^{+01}$ | $8.362 \times 10^{+01}$ | 1.236 |
| 0.020 | $1.108 \times 10^{-26}$ | $2.200 \times 10^{-26}$ | $4.357 \times 10^{-26}$ | 1.990 | 1.500 | $9.642 \times 10^{+01}$ | $1.184 \times 10^{+02}$ | $1.496 \times 10^{+02}$ | 1.257 |
| 0.025 | $3.174 \times 10^{-24}$ | $6.303 \times 10^{-24}$ | $1.248 \times 10^{-23}$ | 1.990 | 1.750 | $1.404 \times 10^{+02}$ | $1.739 \times 10^{+02}$ | $2.234 \times 10^{+02}$ | 1.274 |
| 0.030 | $2.366 \times 10^{-22}$ | $4.699 \times 10^{-22}$ | $9.304 \times 10^{-22}$ | 1.990 | 2.000 | $1.838 \times 10^{+02}$ | $2.295 \times 10^{+02}$ | $2.984 \times 10^{+02}$ | 1.287 |
| 0.040 | $1.257 \times 10^{-19}$ | $2.497 \times 10^{-19}$ | $4.944 \times 10^{-19}$ | 1.990 | 2.500 | $2.624 \times 10^{+02}$ | $3.313 \times 10^{+02}$ | $4.377 \times 10^{+02}$ | 1.304 |
| 0.050 | $1.083 \times 10^{-17}$ | $2.151 \times 10^{-17}$ | $4.258 \times 10^{-17}$ | 1.990 | 3.000 | $3.277 \times 10^{+02}$ | $4.171 \times 10^{+02}$ | $5.557 \times 10^{+02}$ | 1.313 |
| 0.060 | $3.224 \times 10^{-16}$ | $6.405 \times 10^{-16}$ | $1.268 \times 10^{-15}$ | 1.990 | 3.500 | $3.825 \times 10^{+02}$ | $4.909 \times 10^{+02}$ | $6.565 \times 10^{+02}$ | 1.319 |
| 0.070 | $4.831 \times 10^{-15}$ | $9.598 \times 10^{-15}$ | $1.900 \times 10^{-14}$ | 1.990 | 4.000 | $4.301 \times 10^{+02}$ | $5.574 \times 10^{+02}$ | $7.487 \times 10^{+02}$ | 1.326 |
| 0.080 | $4.496 \times 10^{-14}$ | $8.932 \times 10^{-14}$ | $1.768 \times 10^{-13}$ | 1.990 | 5.000 | $5.150 \times 10^{+02}$ | $6.837 \times 10^{+02}$ | $9.396 \times 10^{+02}$ | 1.356 |
| 0.090 | $2.956 \times 10^{-13}$ | $5.873 \times 10^{-13}$ | $1.163 \times 10^{-12}$ | 1.990 | 6.000 | $5.940 \times 10^{+02}$ | $8.158 \times 10^{+02}$ | $1.165 \times 10^{+03}$ | 1.409 |
| 0.100 | $1.494 \times 10^{-12}$ | $2.968 \times 10^{-12}$ | $5.876 \times 10^{-12}$ | 1.990 | 7.000 | $6.704 \times 10^{+02}$ | $9.546 \times 10^{+02}$ | $1.430 \times 10^{+03}$ | 1.473 |
| 0.110 | $6.151 \times 10^{-12}$ | $1.222 \times 10^{-11}$ | $2.419 \times 10^{-11}$ | 1.990 | 8.000 | $7.442 \times 10^{+02}$ | $1.099 \times 10^{+03}$ | $1.725 \times 10^{+03}$ | 1.536 |
| 0.120 | $2.149 \times 10^{-11}$ | $4.270 \times 10^{-11}$ | $8.451 \times 10^{-11}$ | 1.990 | 9.000 | $8.132 \times 10^{+02}$ | $1.242 \times 10^{+03}$ | $2.030 \times 10^{+03}$ | 1.592 |
| 0.130 | $6.574 \times 10^{-11}$ | $1.306 \times 10^{-10}$ | $2.584 \times 10^{-10}$ | 1.990 | 10.000 | $8.775 \times 10^{+02}$ | $1.380 \times 10^{+03}$ | $2.330 \times 10^{+03}$ | 1.640 |

**Supplemental Material**

Full output of the RatesMC calculations of the $^{23}$Mg$(p,\gamma)^{24}$Al reaction rate, incorporating the present results for the $E_{cm} \approx 480$ keV resonance.

```
23Mg(p,g)24Al
Calculated with RatesMC 2.3.0 (git hash:9395b52b05) on Wed Jul  8 10:09:16 2026
Samples = 50000
 T9      RRate_2low      RRate_low       Classical Rate  Median Rate     Mean Rate
RRate_high     RRate_2high     Log-Normal mu      Log-Normal sigma A-D Statistic
 0.001   1.846e-86       3.580e-86       7.153e-86       7.148e-86       9.063e-86
1.418e-85       2.772e-85       -1.961e+02          6.922e-01       2.598e-01
 0.002   6.047e-67       1.173e-66       2.344e-66       2.342e-66       2.969e-66
4.648e-66       9.084e-66       -1.511e+02          6.922e-01       2.598e-01
 0.003   1.792e-57       3.475e-57       6.944e-57       6.939e-57       8.798e-57
1.377e-56       2.691e-56       -1.293e+02          6.922e-01       2.598e-01
 0.004   1.693e-51       3.285e-51       6.563e-51       6.558e-51       8.315e-51
1.301e-50       2.544e-50       -1.156e+02          6.922e-01       2.598e-01
 0.005   3.015e-47       5.849e-47       1.169e-46       1.168e-46       1.481e-46
2.317e-46       4.529e-46       -1.058e+02          6.922e-01       2.598e-01
 0.006   5.267e-44       1.022e-43       2.041e-43       2.040e-43       2.586e-43
4.048e-43       7.911e-43       -9.830e+01          6.922e-01       2.598e-01
 0.007   2.082e-41       4.006e-41       7.949e-41       7.956e-41       1.008e-40
1.575e-40       3.077e-40       -9.233e+01          6.880e-01       2.534e-01
 0.008   2.846e-39       5.477e-39       1.087e-38       1.088e-38       1.378e-38
2.154e-38       4.207e-38       -8.741e+01          6.880e-01       2.533e-01
 0.009   1.818e-37       3.497e-37       6.939e-37       6.945e-37       8.799e-37
1.375e-36       2.686e-36       -8.326e+01          6.881e-01       2.531e-01
 0.010   6.518e-36       1.254e-35       2.488e-35       2.490e-35       3.155e-35
4.932e-35       9.633e-35       -7.968e+01          6.881e-01       2.530e-01
 0.011   1.490e-34       2.867e-34       5.688e-34       5.693e-34       7.213e-34
1.127e-33       2.202e-33       -7.655e+01          6.881e-01       2.529e-01
 0.012   2.377e-33       4.573e-33       9.073e-33       9.081e-33       1.151e-32
1.798e-32       3.513e-32       -7.378e+01          6.881e-01       2.528e-01
 0.013   2.828e-32       5.440e-32       1.079e-31       1.080e-31       1.369e-31
2.139e-31       4.179e-31       -7.130e+01          6.881e-01       2.527e-01
 0.014   2.638e-31       5.075e-31       1.007e-30       1.008e-30       1.277e-30
1.996e-30       3.898e-30       -6.907e+01          6.881e-01       2.526e-01
 0.015   2.006e-30       3.859e-30       7.658e-30       7.664e-30       9.710e-30
1.518e-29       2.964e-29       -6.704e+01          6.881e-01       2.525e-01
 0.016   1.282e-29       2.467e-29       4.895e-29       4.899e-29       6.207e-29
9.701e-29       1.895e-28       -6.519e+01          6.881e-01       2.524e-01
 0.018   3.413e-28       6.567e-28       1.303e-27       1.304e-27       1.652e-27
2.582e-27       5.044e-27       -6.190e+01          6.881e-01       2.522e-01
 0.020   5.758e-27       1.108e-26       2.198e-26       2.200e-26       2.787e-26
4.357e-26       8.509e-26       -5.908e+01          6.881e-01       2.521e-01
 0.025   1.650e-24       3.174e-24       6.298e-24       6.303e-24       7.986e-24
1.248e-23       2.438e-23       -5.342e+01          6.881e-01       2.518e-01
 0.030   1.230e-22       2.366e-22       4.695e-22       4.699e-22       5.953e-22
9.304e-22       1.817e-21       -4.911e+01          6.881e-01       2.515e-01
 0.040   6.534e-20       1.257e-19       2.494e-19       2.497e-19       3.163e-19
4.944e-19       9.654e-19       -4.283e+01          6.882e-01       2.512e-01
 0.050   5.628e-18       1.083e-17       2.148e-17       2.151e-17       2.724e-17
4.258e-17       8.314e-17       -3.838e+01          6.882e-01       2.511e-01
 0.060   1.676e-16       3.224e-16       6.399e-16       6.405e-16       8.114e-16
1.268e-15       2.476e-15       -3.499e+01          6.882e-01       2.510e-01
 0.070   2.511e-15       4.831e-15       9.588e-15       9.598e-15       1.216e-14
1.900e-14       3.710e-14       -3.228e+01          6.882e-01       2.509e-01
 0.080   2.336e-14       4.496e-14       8.922e-14       8.932e-14       1.131e-13
1.768e-13       3.453e-13       -3.005e+01          6.882e-01       2.509e-01
```

| | | | | | | | | | | |
|---|---|---|---|---|---|---|---|---|---|---|
| 0.090 | 1.536e-13 | 2.956e-13 | 5.866e-13 | 5.873e-13 | 7.439e-13 | 1.163e-12 | 2.270e-12 | -2.816e+01 | 6.882e-01 | 2.510e-01 |
| 0.100 | 7.765e-13 | 1.494e-12 | 2.965e-12 | 2.968e-12 | 3.760e-12 | 5.876e-12 | 1.147e-11 | -2.654e+01 | 6.882e-01 | 2.511e-01 |
| 0.110 | 3.197e-12 | 6.151e-12 | 1.220e-11 | 1.222e-11 | 1.548e-11 | 2.419e-11 | 4.723e-11 | -2.513e+01 | 6.881e-01 | 2.514e-01 |
| 0.120 | 1.117e-11 | 2.149e-11 | 4.264e-11 | 4.270e-11 | 5.408e-11 | 8.451e-11 | 1.650e-10 | -2.388e+01 | 6.881e-01 | 2.519e-01 |
| 0.130 | 3.417e-11 | 6.574e-11 | 1.304e-10 | 1.306e-10 | 1.653e-10 | 2.584e-10 | 5.045e-10 | -2.276e+01 | 6.879e-01 | 2.550e-01 |
| 0.140 | 9.399e-11 | 1.804e-10 | 3.573e-10 | 3.578e-10 | 4.530e-10 | 7.076e-10 | 1.381e-09 | -2.175e+01 | 6.867e-01 | 2.808e-01 |
| 0.150 | 2.404e-10 | 4.556e-10 | 8.969e-10 | 8.980e-10 | 1.135e-09 | 1.771e-09 | 3.451e-09 | -2.083e+01 | 6.813e-01 | 5.683e-01 |
| 0.160 | 6.044e-10 | 1.099e-09 | 2.116e-09 | 2.118e-09 | 2.667e-09 | 4.130e-09 | 8.001e-09 | -1.997e+01 | 6.620e-01 | 3.708e+00 |
| 0.180 | 4.892e-09 | 7.165e-09 | 1.163e-08 | 1.168e-08 | 1.405e-08 | 2.046e-08 | 3.743e-08 | -1.823e+01 | 5.244e-01 | 1.057e+02 |
| 0.200 | 5.127e-08 | 6.452e-08 | 8.259e-08 | 8.412e-08 | 9.141e-08 | 1.164e-07 | 1.763e-07 | -1.626e+01 | 3.128e-01 | 2.231e+02 |
| 0.250 | 6.900e-06 | 8.324e-06 | 1.010e-05 | 1.006e-05 | 1.025e-05 | 1.218e-05 | 1.462e-05 | -1.151e+01 | 1.915e-01 | 1.772e-01 |
| 0.300 | 2.103e-04 | 2.528e-04 | 3.091e-04 | 3.051e-04 | 3.109e-04 | 3.690e-04 | 4.436e-04 | -8.094e+00 | 1.906e-01 | 2.272e-01 |
| 0.350 | 2.389e-03 | 2.865e-03 | 3.501e-03 | 3.453e-03 | 3.516e-03 | 4.164e-03 | 4.994e-03 | -5.668e+00 | 1.885e-01 | 1.916e-01 |
| 0.400 | 1.450e-02 | 1.735e-02 | 2.114e-02 | 2.087e-02 | 2.124e-02 | 2.510e-02 | 3.004e-02 | -3.869e+00 | 1.862e-01 | 1.809e-01 |
| 0.450 | 5.806e-02 | 6.933e-02 | 8.415e-02 | 8.319e-02 | 8.461e-02 | 9.984e-02 | 1.193e-01 | -2.487e+00 | 1.839e-01 | 1.799e-01 |
| 0.500 | 1.739e-01 | 2.074e-01 | 2.508e-01 | 2.484e-01 | 2.526e-01 | 2.976e-01 | 3.548e-01 | -1.393e+00 | 1.820e-01 | 1.978e-01 |
| 0.600 | 8.802e-01 | 1.048e+00 | 1.257e+00 | 1.252e+00 | 1.272e+00 | 1.495e+00 | 1.777e+00 | 2.245e-01 | 1.792e-01 | 2.879e-01 |
| 0.700 | 2.733e+00 | 3.254e+00 | 3.886e+00 | 3.894e+00 | 3.953e+00 | 4.644e+00 | 5.531e+00 | 1.358e+00 | 1.791e-01 | 3.827e-01 |
| 0.800 | 6.285e+00 | 7.496e+00 | 8.918e+00 | 8.981e+00 | 9.127e+00 | 1.074e+01 | 1.286e+01 | 2.195e+00 | 1.818e-01 | 4.995e-01 |
| 0.900 | 1.183e+01 | 1.415e+01 | 1.682e+01 | 1.700e+01 | 1.733e+01 | 2.045e+01 | 2.475e+01 | 2.835e+00 | 1.869e-01 | 2.940e+00 |
| 1.000 | 1.945e+01 | 2.330e+01 | 2.772e+01 | 2.807e+01 | 2.872e+01 | 3.403e+01 | 4.173e+01 | 3.339e+00 | 1.936e-01 | 1.063e+01 |
| 1.250 | 4.622e+01 | 5.552e+01 | 6.637e+01 | 6.762e+01 | 6.982e+01 | 8.362e+01 | 1.068e+02 | 4.223e+00 | 2.122e-01 | 5.142e+01 |
| 1.500 | 7.999e+01 | 9.642e+01 | 1.159e+02 | 1.184e+02 | 1.235e+02 | 1.496e+02 | 1.976e+02 | 4.789e+00 | 2.290e-01 | 9.856e+01 |
| 1.750 | 1.158e+02 | 1.404e+02 | 1.695e+02 | 1.739e+02 | 1.828e+02 | 2.234e+02 | 3.013e+02 | 5.178e+00 | 2.423e-01 | 1.325e+02 |
| 2.000 | 1.512e+02 | 1.838e+02 | 2.226e+02 | 2.295e+02 | 2.423e+02 | 2.984e+02 | 4.092e+02 | 5.457e+00 | 2.525e-01 | 1.510e+02 |
| 2.500 | 2.144e+02 | 2.624e+02 | 3.177e+02 | 3.313e+02 | 3.517e+02 | 4.377e+02 | 6.088e+02 | 5.826e+00 | 2.654e-01 | 1.567e+02 |
| 3.000 | 2.659e+02 | 3.277e+02 | 3.952e+02 | 4.171e+02 | 4.435e+02 | 5.557e+02 | 7.741e+02 | 6.056e+00 | 2.723e-01 | 1.392e+02 |
| 3.500 | 3.078e+02 | 3.825e+02 | 4.588e+02 | 4.909e+02 | 5.212e+02 | 6.565e+02 | 9.100e+02 | 6.216e+00 | 2.767e-01 | 1.109e+02 |
| 4.000 | 3.431e+02 | 4.301e+02 | 5.145e+02 | 5.574e+02 | 5.912e+02 | 7.487e+02 | 1.035e+03 | 6.341e+00 | 2.820e-01 | 8.318e+01 |
| 5.000 | 4.005e+02 | 5.150e+02 | 6.219e+02 | 6.837e+02 | 7.296e+02 | 9.396e+02 | 1.321e+03 | 6.545e+00 | 3.042e-01 | 5.994e+01 |

| | | | | | |
|---|---|---|---|---|---|
| 6.000 | 4.512e+02 | 5.940e+02 | 7.412e+02 | 8.158e+02 | 8.851e+02 |
| 1.165e+03 | 1.736e+03 | 6.724e+00 | 3.429e-01 | 7.458e+01 | |
| 7.000 | 4.944e+02 | 6.704e+02 | 8.766e+02 | 9.546e+02 | 1.061e+03 |
| 1.430e+03 | 2.263e+03 | 6.888e+00 | 3.876e-01 | 9.679e+01 | |
| 8.000 | 5.334e+02 | 7.442e+02 | 1.022e+03 | 1.099e+03 | 1.250e+03 |
| 1.725e+03 | 2.874e+03 | 7.033e+00 | 4.295e-01 | 1.057e+02 | |
| 9.000 | 5.648e+02 | 8.132e+02 | 1.170e+03 | 1.242e+03 | 1.441e+03 |
| 2.030e+03 | 3.511e+03 | 7.158e+00 | 4.652e-01 | 1.013e+02 | |
| 10.000 | 5.942e+02 | 8.775e+02 | 1.312e+03 | 1.380e+03 | 1.625e+03 |
| 2.330e+03 | 4.120e+03 | 7.263e+00 | 4.944e-01 | 9.069e+01 | |